\def\editmode{0}

\def\bibfilenames{bibman_refs} 

\def\singlenarrowcol{0}
\if\singlenarrowcol0
    \documentclass[journal]{IEEEtran}
    \IEEEoverridecommandlockouts
\else
    \documentclass[onecolumn]{IEEEtran}
    \usepackage[pass]{geometry}
    \newgeometry{textwidth=252pt, textheight=672pt}
    
\fi

\newenvironment{changes}{\color{blue}}{\ignorespacesafterend}

\usepackage{savesym} %% Needed for compatibility with IEEEtran
\savesymbol{labelindent}
\usepackage{enumitem} % do not include the package enumerate. See below.
\if\editmode1  %%%%%%%%%%%%%%%%%%%%%%%%%%%%%%%%%%%%%%%%%%%%%%%%%%%%%%%%%%%
\DeclareFieldFormat{labelalpha}{\thefield{entrykey}}
\DeclareFieldFormat{extraalpha}{}
\bibliography{\bibfilenames}
\newcommand{\cmt}[1]{\noindent\textcolor{darkgreen}{\underline{[#1]}}} % comment
\newcommand{\hc}[1]{\textcolor{blue}{#1}} % highlight command --> to
\setlistdepth{9}
\newlist{bulletlist}{enumerate}{9}
\setlist[bulletlist,1]{label=$\bullet$}
\setlist[bulletlist,2]{label=$\diamond$}
\setlist[bulletlist,3]{label=$\rightarrow$}
\setlist[bulletlist,4]{label=$\circ$}
\setlist[bulletlist,5]{label=$\square$}
\setlist[bulletlist,6]{label=$\star$}
\setlist[bulletlist,7]{label=$\checkmark$}
\setlist[bulletlist,8]{label=$\Delta$}
\setlist[bulletlist,9]{label=$-$}
\newenvironment{bullets}{\begin{bulletlist}}{\end{bulletlist}}

\newcommand{\blt}[1][noargpassed]{% add a label if an optional argument is passed
  \item%
  \ifthenelse{\equal{#1}{noargpassed}}{}{\cmt{#1}}%
}

\else
\usepackage{cite}
\newcommand{\cmt}[1]{} % comment
\newcommand{\hc}[1]{\textcolor{black}{#1}} % highlight command -->
\newenvironment{bullets}{}{}
\newcommand{\blt}[1][noargpassed]{\ignorespaces}

\fi
\newcommand{\printmybibliography}{
\if\editmode1 
\printbibliography
\else
\bibliography{\bibfilenames}
\fi
}

\usepackage{fixltx2e}

\usepackage{graphicx}

\usepackage{subcaption}              %subfigures with \mbox and \subfigure[]

\usepackage[utf8]{inputenc}

\usepackage{amsfonts}
\usepackage{amsmath}
\usepackage{mathtools}
\usepackage{amssymb}

\usepackage{bm}

\usepackage{color,verbatim}
\usepackage{multirow}
\usepackage{accents}

\usepackage{theoremref}
\usepackage{url}

\makeatletter
\newtoks\@soltoks
\newcommand\printquizsolutions{%
  \the\@soltoks}
\newcommand\quizsolution[1]{%
  \begin{flushright}
    See the solution \hyperref[{qxsol:\themyquiz}]{here}.
  \end{flushright}
  \edef\@soltemp{%
    \the\@soltoks % the previously stored solutions
    \noexpand\solutionpreamble{\themyquiz}
    \unexpanded{
      #1
    }%
  }
  \global\@soltoks=\expandafter{\@soltemp}
}
\newcommand{\solutionpreamble}[1]{%
  \par\medskip\noindent\section{Solution to Quiz~\ref{qx:#1}}\label{qxsol:#1}\enspace\ignorespaces
}
\makeatother

\newcounter{rulecounter}
\newcommand{\resetrule}{ \setcounter{rulecounter}{0}}
\resetrule

\newtheorem{myauxproblem}{Problem}
\newtheorem{myauxoptionalproblem}{Optional Problem}
\newsavebox{\selvestebox}
\newenvironment{colbox}[1]
  {\newcommand\colboxcolor{#1}%
   \begin{lrbox}{\selvestebox}%
   \begin{minipage}{\linewidth}}
  {\end{minipage}\end{lrbox}%
   \begin{center}
   \colorbox{\colboxcolor}{\usebox{\selvestebox}}
   \end{center}}

\definecolor{orange}{rgb}{1,0.8,0}
\definecolor{gray}{rgb}{.9,0.9,0.9}
\definecolor{darkgray}{rgb}{.3,0.3,0.3}
\definecolor{darkblue}{rgb}{.1,0.0,0.3}
\definecolor{lightblue}{rgb}{0.7,0.7,1}
\definecolor{lightred}{rgb}{1,0.7,.7}
\definecolor{purple}{RGB}{204,153,255}
\definecolor{lightgray}{rgb}{.95,0.95,0.95}
\definecolor{lightgreen}{rgb}{0.6,0.8,0.6}
\definecolor{darkgreen}{rgb}{0.05,0.3,0.05}
\definecolor{pistachio}{RGB}{204,255,153}
\definecolor{paleturquoise}{RGB}{175,238,238}
\definecolor{yellow}{RGB}{255,255,153}

\newcommand{\ra}{$\rightarrow$~}

\newcommand{\ubar}[1]{\underaccent{\bar}{#1}}

\newcommand{\brackets}[1]{\left\{#1\right\}}

\newcommand{\dbar}[1]{{\bar{\bar{ #1}}}}

\newcommand{\hbm}[1]{{\hat{\bm #1}}}
\newcommand{\inv}{^{-1}}

\newcommand{\ffield}{\mathbb{F}}
\newcommand{\cfield}{\mathbb{C}}
\newcommand{\rfield}{\mathbb{R}}

\newcommand{\cov}{\mathop{\rm Cov}}

\newcommand{\spanv}{\mathop{\rm span}}

\newcommand{\transpose}{^\top}
 \newcommand{\define}{:=}
\newcommand{\expected}{\mathop{\textrm{E}}\nolimits}

\newcommand{\minimize}{\mathop{\text{minimize}}}

\DeclareMathOperator*{\argmin}{arg\,min}
\DeclareMathOperator*{\argmax}{arg\,max}

\newtheorem{myproposition}{Proposition}
\newtheorem{myquestion}{Question}
\newtheorem{myquiz}{Quiz}
\newtheorem{myremark}{Remark}
\newtheorem{myproblemstatement}{Problem}
\newtheorem{mylemma}{Lemma}
\newtheorem{mytheorem}{Theorem}
\newtheorem{mydefinition}{Definition}
\newtheorem{mycorollary}{Corollary}
\newtheorem{myexample}{Example}

\renewcommand{\expected}{\hc{\mathbb{E}}} % 

\newcommand{\area}{\hc{\mathcal{X}}} % geographical area
\newcommand{\areaside}{\hc{L}} % 
\newcommand{\areasidex}{\hc{L}_x} % 
\newcommand{\areasidey}{\hc{L}_y} % 
\newcommand{\lgarea}{\hc{\underline{\mathcal{X}}}} % large geographical area
\newcommand{\lgareasidex}{\hc{\underbar{L}}_x} % 
\newcommand{\lgareasidey}{\hc{\underbar{L}}_y} % 

\newcommand{\measnum}{\hc{N}}

\newcommand{\lgmeasnum}{\hc{\underbar{N}}}
\newcommand{\loc}{{\hc{{\bm{x}}}}} % measurement location
\newcommand{\measnoise}{\hc{z}} %  measurement noise
\newcommand{\measind}{{\hc{n}}}
\newcommand{\measset}{{\hc{\mathcal{M}}}}

\newcommand{\obsnum}{{\hc{\measnum_{\text{obs}}}}}

\newcommand{\measindset}{{\hc{\mathcal{N}}}}
\newcommand{\obsmeasindset}{{\hc{\measindset_{\text{obs}}}}}
\newcommand{\nobsmeasindset}{{\hc{\measindset_{\text{nobs}}}}}

\newcommand{\txpower}{\hc{P_{\text{Tx}}}}
\newcommand{\antgain}{\hc{G}}
\newcommand{\loss}{\hc{s}}

\newcommand{\pow}{{\hc{\gamma}}} % 
\newcommand{\powmeas}[1]{\hc{\tilde{\pow}}_{#1}} % 
\newcommand{\powest}{{\hc{\hat \pow}}} % estimated power gain
\newcommand{\powmat}{\hc{\bm{\Gamma}}}
\newcommand{\powmeasvec}{\hc{\tilde{\bm{\pow}}}}
\newcommand{\powmeasmat}{\hc{\tilde{\powmat}}} % 
\newcommand{\pownff}{{\hc{\gamma}_{\text{NSSF}}}} % power no fast fading
\newcommand{\powestmat}{\hc{\hbm{\Gamma}}}

\newcommand{\expectation}{{\hc{\mathbb{E}}}} % posterior locatiion

\newcommand{\grid}{{\hc{\mathcal{G}}}} % dnn function
\newcommand{\gridnobs}{{\hc{\text{G-nobs}}}} %
\newcommand{\gridall}{{\hc{\text{G-all}}}} % 

\newcommand{\neighbornum}{\hc{K}} % 
\newcommand{\neighborind}{{\hc{k}}} % 
\newcommand{\neighborindfun}{\hc{\nu}} % 

\newcommand{\powmean}{\hc{\mu_\pow}} % power mean

\newcommand{\krrcoef}{\hc{\alpha}} % 
\newcommand{\kernelfun}{\hc{\kappa}} % 
\newcommand{\regpar}{\hc{\rho}} %
\newcommand{\kernelwidth}{\hc{s}} % 

\newcommand{\samplingmask}{\hc{\bm{S}}} % 

\newcommand{\gridx}{\hc{N_x}}  % grid num points in x-axis
\newcommand{\gridy}{\hc{N_y}}   % grid num points in y-axis

\newcommand{\lggridx}{\hc{\underline{N}}_x}  %lg grid num points in x-axis
\newcommand{\lggridy}{\hc{\underline{N}_y}}   % lg grid num points in y-axis

\newcommand{\shad}{{\hc{ s}}} % zero-mean shadowing
\newcommand{\dist}{\hc{\delta}_{\shad}}
\newcommand{\ushadvar}{{\hc{\sigma^2_{\hc{ \shad}}}}} % unnormalized shadowing variance
\newcommand{\gridspacing}{\hc{\Delta}}  % grid spacing

 \newcommand{\journal}[1]{#1} % material to omit from the conference
\newcommand{\conference}[1]{} % material to omit from the journal paper

\newcommand{\ifroom}[1]{}

\newcommand{\nextversion}[1]{}

\begin{document}

%%%%%%%%%%%%%%%%%%%%%%%%%%%%%%%%%%%%%%%%%%%%%%%%%%%%%%%%%%%%%%%%%%%%%
\title{Radio Map Estimation: \\Empirical Validation and Analysis}

\author{\IEEEauthorblockN{Raju Shrestha, Tien Ngoc Ha, Pham Q. Viet,
        and Daniel Romero}\\
    \IEEEauthorblockA{{Department of Information and Communication Technology},
        {University of Agder},
        Norway\\
        \texttt{\{raju.shrestha,tien.n.ha,viet.q.pham,daniel.romero\}@uia.no}}
}

\maketitle
%%%%%%%%%%%%%%%%%%%%%%%%%%%%%%%%%%%%%%%%%%%%%%%%%%%%%%%%%%%%%%%%%%%%%

%\renewcommand*{\thefootnote}{}

\begin{abstract}
    Radio maps provide metrics such as the received signal strength at every
    location in a geographical region of interest. Extensive research has been
    carried out in this context, but it relies almost exclusively on
    synthetic-data experiments. Thus, the practical aspects of the radio map
    estimation (RME) problem as well as  the performance of existing estimators
    in the real world remain unknown. To fill this gap end, this paper  puts
    forth the first comprehensive, rigorous, and reproducible study of RME with
    real data. The main contributions include (C1) an assessment of the
    viability of RME based on the estimation error that can be achieved, (C2)
    the analysis of the main phenomena and trade-offs involved in RME, including
    the experimental verification of theoretical findings in the literature, and
    (C3) a thorough evaluation of a wide range of estimators  on real-world
    data. Remarkably, this  reveals that the performance gain of existing deep
    estimators in their pure form may not compensate for their complexity. A
    simple enhancement (C4) is proposed to alleviate this issue. The vast amount
    of data collected for this study is published along with the developed
    simulator to enable research on new schemes, hopefully bringing RME one step
    closer to  practical deployment.
    % whether radio map estimation (RME) can attain a sufficiently become
    % someday a real-world technology and (ii) which are the most effective
    % estimation approaches. To fill this gap, this paper presents the first
    % comprehensive, rigorous, and reproducible study of RME with real data.
    % General aspects of the RME problem are investigated and the capabilities of
    % existing estimators are assessed and compared using two large-scale
    % measurement datasets collected in this work. Recent theoretical findings in
    % RME are empirically corroborated and a large collection of observations are
    % presented to guide researchers and practitioners. A surprising observation
    % is that estimators based on convolutional neural networks (CNN) do not
    % perform significantly better than more traditional interpolation methods.
    % This suggests that future research should discontinue pure-CNN estimators in 
    % favor of hybrid algorithms that combine CNN and traditional methods. To
    % corroborate this observation, one of the simplest algorithms that one can
    % conceive in this family is tested and found to outperform all existing
    % schemes.    

\end{abstract}

\begin{keywords}
    Radio environment map, radio knowledge map, radio map estimation, unmanned aerial
    vehicles.\footnote{This research has been funded in part by the Research
        Council of Norway under IKTPLUSS grant 311994.}
\end{keywords}

\IEEEpeerreviewmaketitle

\section{Introduction}
\label{sec:intro}

\begin{bullets}
    \blt[motivation]Radio maps provide a radio frequency (RF) metric at each spatial location of a geographical region~\cite{romero2022cartography,alayafeki2008cartography}. Examples of such metrics include the received signal power, interference power, power spectral density (PSD), electromagnetic absorption, and channel gain. Fig.~\ref{fig:3D_true_map} depicts an example of a power map. Due to  their ability to convey information about signal propagation, interference sources, and channel occupancy, radio maps are crucial in many applications, including cellular communications, device-to-device communications, network planning, frequency planning,  path planning for autonomous vehicles, dynamic spectrum access, aerial traffic management in unmanned aerial systems, and fingerprinting localization~\cite{romero2022cartography}. Radio maps are typically  estimated by interpolating measurements acquired across the region of~interest.

    \begin{figure}[t]
        \centering
        \includegraphics[width=1\linewidth]{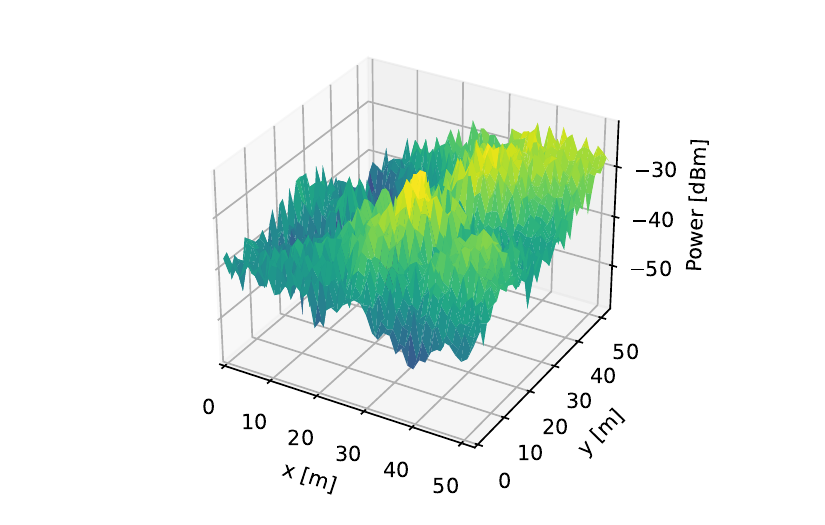}
        \caption{Power map constructed via
            grid discretization (Sec.~\ref{sec:gridawarerme}) of one of the collected measurement sets.}
        \label{fig:3D_true_map}
    \end{figure}

    \blt[SoA/literature\ra
        limitations] Most works on radio map estimation  (RME) adhere to a common profile: they propose an estimator and compare it with a small number of benchmarks on \emph{synthetic} data; see~\cite{hoyhtya2016survey,romero2022cartography\journal{,pesko2014survey}}
    for lists of references. As a result, important open questions remain. For example:
    \begin{bullets}
        \blt (Q1) Is the estimation error of contemporary algorithms sufficiently low to render RME a viable technology in practical applications?
        \blt (Q2) What are the main trade-offs and features   of the RME problem in the real world?
        \blt (Q3) Which algorithms and families of algorithms work best in practice?
        \blt (Q4) Do deep learning algorithms yield practical benefits that compensate for their high complexity and data requirements? Note that no past work managed to effectively train deep learning estimators on real data.
    \end{bullets}
    % \begin{bullets}%
    %     \blt[RME problem]On the one hand, the RME problem itself is not well
    %     understood. Although  recent  theoretical results offer valuable
    %     insight~\cite{romero2024theoretical}, understanding certain aspects of the problem require an empirical approach.
    %     %
    %     \blt[Estimators]On the other hand, it is unclear how existing estimators perform and compare in practice. As detailed in Sec.~\ref{sec:relatedwork}, the work in this context is still highly limited.
    % \end{bullets}%

    \blt[contributions]These questions motivate this paper, which is \emph{the first comprehensive, rigorous, and reproducible study of RME with real data}. Each main contribution targets one of these questions:\footnote{Relative to the conference version~\cite{shrestha2023empirical}, this paper (i) considers two more real datasets (the LOS USRP and the  4G datasets), (ii) conducts comprehensive empirical validation with additional benchmarks, including trained traditional estimators, (iii) introduces  hybrid estimators, (iv) quantifies the required training data, (v) and includes much more detailed analysis. }
    \begin{enumerate}[label=C\arabic*)]%

        \item\cmt{Viability}The \emph{practical viability} of RME is assessed by quantifying the estimation error of a broad set of estimators in different propagation scenarios and by determining the number of observations required to attain a target accuracy.
        \item\cmt{analysis of the RME problem}The  main phenomena involved in the RME problem are empirically  investigated.
        \begin{bullets}%
            \blt[small-scale
                fading.] Remarkably, the impact of small-scale fading is analyzed and approaches to mitigate its effect are explored.
            \blt[source proximity]In addition, the theoretical
            result in~\cite{romero2024theoretical} that establishes that  the complexity of
            the RME problem decreases with the distance to  the transmitters is verified.
        \end{bullets}%
        See Sec.~\ref{sec:keyobservations} for a list of the main findings.

        \item\cmt{Comparison of estimators}The main  estimators  in the literature, including  traditional estimators and estimators based on convolutional neural networks (CNNs), 
         are extensively analyzed. The main observations are condensed in Table~\ref{tab:summary}.

        \item\cmt{Proposed estimator family}
         CNN estimators are seen  not to perform as well as expected relative to traditional methods. A key reason is that they approach the RME problem as an image processing problem instead of exploiting the fact that a radio map is a spatial field with a significant \emph{parsimonious}\footnote{In this context, \emph{parsimonious} means parameterizable with a small number of parameters.} \emph{component}. To verify this hypothesis, a simple \emph{hybrid estimator} that combines CNN and traditional estimators is designed and tested. The resulting scheme outperforms both traditional and CNN estimators, which opens the door for research on  more sophisticated hybrid algorithms.

        % This suggests that it is convenient to discontinue research on pure-CNN estimators in favor of estimators that combine CNNs with traditional approaches. To verify this, a very simple estimator of this kind is developed and tested. Despite its simplicity, this scheme sets the state of the art, which calls for research on more sophisticated estimators in this family.

        % \item\cmt{simulator}\begin{changes}An open-source simulator is  released to allow the community to test and compare other estimators on real data.
        % \end{changes}
    \end{enumerate}

    \cmt{dataset}The vast amount of data necessary to carry out this study while ensuring statistical significance  required \emph{the greatest data collection effort  by far  ever undertaken in RME}. To make it possible in a feasible time frame, a novel data collection system using software defined radios onboard unmanned aerial vehicles (UAVs) was designed and implemented.
    The
    resulting datasets are released together with this paper  and constitute \emph{the first public RME datasets} with real data. They will hopefully assist the research community in comparing and developing new estimators.
    \cmt{first
        large-scale public dataset}The datasets here are also the first ones of a sufficient size to train and test CNN-based estimators on real data.
    % \begin{bullets}%
    %     \blt[USRP data collection]One dataset comprises received signal strength (RSS) measurements at 320,000 locations with a custom-made system based on a software-defined radio (SDR) on board an unmanned aerial vehicle (UAV).
    %     %
    %     % To collect USRP data, a measurement system was developed utilizing an unmanned aerial vehicle (UAV) equipped with a software-defined radio. The UAV follows a predetermined trajectory over the targeted area, capturing various channel configurations. Accurate positioning is achieved by integrating measurements from an on-board inertial measurement unit (IMU) with differentially obtained GPS data.
    %     %
    %     \blt[4G data collection]The second dataset comprises three metrics of several LTE cells collected at 4,095 locations by an LTE modem on board a UAV.
    %     %           
    % \end{bullets}        
    \cmt{measurements}Measurements were collected at around 324,000 locations, which is \emph{2,000 times more} than existing works where RME data was collected with UAVs; cf. Sec.~\ref{sec:relatedwork}.

    \blt[error reduction]Several measures were adopted to reduce measurement error, namely (m1) static and dynamic measurements, (m2) coherent pilot-driven channel estimation, (m3) pilot-only transmissions, (m4)
    accurate geolocation by fusing inertial
    measurements and \emph{real-time kinematics} (RTK) GPS estimates, (m5) constant-yaw trajectories for limiting error due  antenna directivity, (m6) selection of a frequency band without interference, (m7) spatial averaging to counteract small-scale fading, (m8) outlier detection in the time domain to counteract impulsive noise caused by the motors, and (m9) the usage of a single receiver  to avoid calibration errors.

    \blt[paper structure] The rest of the paper is structured as follows. Sec.~\ref{sec:relatedwork} summarizes the related work.
    Sec.~\ref{sec:rme} reviews the RME problem along with some of the most
    popular estimators in the literature. Sec.~\ref{sec:collection} describes
    the data collection system and procedure. The empirical study using this  data
    is then presented  in Sec.~\ref{sec:experiments}. Finally,
    Sec.~\ref{sec:closingremarks} provides a summary of the key findings, limitations, and the main conclusions.

    \emph{Terminology.} Some terms defined and used  throughout the paper are listed in Table~\ref{tab:terms}.
    \begin{table}[]
        \caption{Summary of the main terminology. }
        \label{tab:terms}
        \begin{tabular}{|p{1.8cm}|p{6.2cm}|}
            \hline
            \multicolumn{1}{|c|}{\textbf{Terms}} & \multicolumn{1}{c|}{\textbf{Definition}   }                                                                         \\ \hline
            Dataset                              & Collection of measurement sets.                                                                                     \\ \hline
            Measurement set                      & Collection of $\lgmeasnum$ measurements of a metric corresponding to a given geographical location and transmitter.
            \\ \hline
            Estimation instance                  &
            Subset of $\measnum$ out of the $\lgmeasnum$ measurements in a measurement set obtained by drawing a patch; cf. Sec.~\ref{sec:collection}.
            \\ \hline
            Set of observed measurements         &
            Subset of  $\obsnum$ measurements or grid points obtained from an estimation instance as described in Sec.~\ref{sec:metrics}.
            \\ \hline
        \end{tabular}
    \end{table}

    \emph{Notation.} Bold lowercase (resp. uppercase) denote column vectors (resp. matrices). Calligraphic letters represent sets.

\end{bullets}

\section{Related Work}
\label{sec:relatedwork}

To the best of our knowledge, no work in the literature provides a general empirical study of RME with real data. While providing valuable insights, most works with real data focus on analyzing a specific algorithm rather than  the RME problem itself. As a result, a comprehensive \emph{performance assessment} and \emph{neutral comparison} of algorithms is still missing. In particular, the papers in the literature (i) consider only a small (and not up-to-date) set of algorithms, (ii) try to advocate the algorithm that they propose, which does not incentivize proper training and parameter setting for the benchmarks, (iii) do not share the data sets to reproduce the experiments, (iv)  use datasets where the propagation conditions are not sufficiently heterogeneous,  and (v)  rely on datasets that are not sufficiently large to train estimators based on deep learning and to draw statistically significant conclusions.

% \begin{changes}
%     In all cases, the number of collected measurements is insufficient to train and test CNN-based estimators and to draw general conclusions about the  RME problem. Besides, none of these works compares a representative set of state-of-the-art estimators.
% \end{changes}

\begin{bullets}%
    \blt[Pure RME problems]%

    \begin{bullets}%

        \blt[xiang2005hidden]
        % Other works with real data are more targeted to
        % assess the performance of a specific estimator. 
        The first work to collect real data for RME seems to be
        \cite{xiang2005hidden}, where the RSS of a base station is measured at 700
        indoor locations using several types of antennas to evaluate the
        performance of an algorithm that capitalizes on the transmitter location
        and a propagation model.
        \blt[hu2013efficient] In \cite{hu2013efficient}, the RSS of multiple
        WiFi access points is measured at 337 indoor locations and used to
        compare the performance of the proposed matrix completion algorithm
        against two benchmarks.
        \blt[yang2016compressive] In \cite{yang2016compressive}, the RSS of 4
        WiFi access points is measured at 124 indoor locations to compare
        the proposed  compressed sensing estimator with a basis
        pursuit and basic model-based path-loss benchmarks.
        \blt[niu2018recnet] In \cite{niu2018recnet}, the RSS of several WiFi
        access points is measured at 1035  indoor locations to compare the proposed
        deep learning algorithm with linear regression.
        \blt[elfriakh2018crowdsourced] In \cite{elfriakh2018crowdsourced},  4
        non-deep interpolators are compared using a single
        performance metric on a dataset of 1,661 RSS measurement locations.
        Interestingly, it is observed that the performance of the considered
        estimators does not greatly differ.
        \blt[iwasaki2020transferbasedpower] In
        \cite{iwasaki2020transferbasedpower}, 100 RSS measurements were
        collected  outdoors to compare their algorithm (which requires
        the transmitter location) against Kriging, linear regression, and a
        simulation-based scheme.%
    \end{bullets}

    \blt[UAV+real data+RME]There are also a few works where UAVs collect data for RME~\cite{ivanov2023limited,ivanov2024deep,platzgummer2019coverage}. These works constitute a valuable proof of concept but the number of measurement locations is in all cases smaller than 160.
    \blt[related problems]Some other works (e.g. \cite{parera2020tiltdependent,hayashi2020studyradiopropagation,ostlin2010macrocell,popoola2019determination}) considered related problems and collected real data. 
    \ifroom{%
        \begin{bullets}%
            \blt[parera2020tiltdependent]For example, in
            \cite{parera2020tiltdependent}, 600 \begin{changes} reference signal received power (RSRP) \end{changes} measurements are collected
            outdoors for various tilt values of the antennas of an LTE base
            station with the purpose of predicting RSS at all locations for other
            tilt values.
            \blt[hayashi2020studyradiopropagation] Another related problem is
            considered in \cite{hayashi2020studyradiopropagation}, where path loss
            is predicted using a neural network. Performance is studied with
            measurements collected at 36,000 locations, but no comparison with
            other algorithms is carried out.
            \blt[ostlin2010macrocell] Yet another example, in
            \cite{ostlin2010macrocell}, published before the advent of deep
            learning, 1238 measurements were collected outdoors and fed to
            artificial neural networks to predict path loss.
            \blt[popoola2019determination] The work in
            \cite{popoola2019determination}, which relies on 538 RSS measurements,
            follows similar lines as \cite{ostlin2010macrocell}.
        \end{bullets}
    }

\end{bullets}

\section{Radio Map Estimation}
\label{sec:rme}

Amidst the diverse set of RME formulations, this paper concentrates on the two most
prevalent ones, here referred to as the grid-aware and grid-agnostic
formulations. For simplicity, the exposition  assumes that the signal strength
is quantified using the received signal power, but other metrics will also be
considered afterwards.

\subsection{Model}

\begin{bullets}
    \blt[model]
    \begin{bullets}
        \blt[region] Let $\area\subset\rfield^{2}$ encompass the Cartesian
        coordinates of all points within the region of interest, typically a
        rectangular area in a horizontal plane.
        \blt[map]A \emph{power map} is a function that returns the received signal power
        $\pow(\loc)$ for each location
        $\loc \in \area$. This power is the result of the
        contribution of one or multiple transmitters operating on a given frequency band.
        \blt[example with 1 tx] For example, if there is a
        single transmitter  with transmit power $\txpower$, then
        \begin{align}
            \label{received_power}
            \pow \left( \loc \right) = \txpower + \antgain - \loss^{\text{PL}} (\loc)  - \loss^{\text{S}} (\loc) - \loss^{\text{SSF}} (\loc),
        \end{align}
        where $\loss^{\text{PL}} (\loc), \loss^{\text{S}} (\loc)$,
        $\loss^{\text{SSF}} (\loc)$, and $\antgain$ respectively denote the
        path loss, the loss due to shadowing, the loss due to small-scale
        fading, and the constant gain term that aggregates the gains
        of the antennas and amplifiers.

        \blt[measurement model] The received power is measured by one or
        multiple receivers (or sensors) with isotropic antennas at $\measnum$
        locations $\left\{\loc_{\measind} \right\}_{\measind=1}^{\measnum}
            \subset \area$. The measurement at
        $\loc_{\measind}$ can be written as
        \begin{align}
            \label{eq:measmodel}
            \powmeas{\measind} = \pow \left( \loc_{\measind} \right) + \measnoise_{\measind},
        \end{align}
        where $\measnoise_{\measind}$ represents measurement noise. % The UAV collects a power measurement $\powmeas_{\measind}$ when it is at position $\loc_{\measind} \in \area, \measind=0, 1, \ldots, \measnum-1,$ where $ \measnoise_{\measind} \sim \mathcal{N}(0, \sigma^2)$ models the measurement error and is assumed independent across $\measind$. 

    \end{bullets}

    \subsection{Problem Formulation}
    \label{sec:problem}
    \begin{bullets}
        \blt[motivation \ra small-scale fading]Since the term $\loss^{\text{SSF}}
            (\loc)$ in \eqref{received_power} is caused by multipath, it exhibits a
        spatial variability at a wavelength scale, which for contemporary
        communication systems is not generally greater than tens of centimeters. Since
        accurately estimating $\loss^{\text{SSF}} (\loc)$ would arguably require
        a spacing between measurement locations below the wavelength,  the total number of necessary
        measurement locations would be prohibitive. For this reason, it is common in the RME literature to
        assume that small-scale fading is \emph{averaged out}. To study  the impact of
        such an assumption, which has never been empirically analyzed, the present
        work considers two common problem formulations in the literature, one
        where $\loss^{\text{SSF}} (\loc)$ is accounted for and one where
        $\loss^{\text{SSF}} (\loc)$ is averaged out.
    \end{bullets}

    \subsubsection{Grid-agnostic RME}
    \label{sec:gridagnosticrme}
    \begin{bullets}
        \blt[intro] In this formulation, small-scale fading is not averaged out.
        \blt[formulation]%
        \begin{bullets}%
            \blt[given] Given  $\obsnum$ observed measurements
            $\left\{(\loc_{\measind},
                \powmeas{\measind})\right\}_{\measind=1}^{\obsnum}$,
            \blt[requested]the problem is to estimate  $\pow(\loc)$, $\loc\in\area$.
        \end{bullets}
        \blt[why $\obsnum$]In the operational phase,  $\obsnum$ will be set to $\measnum$. However, for training purposes, it is convenient to split the $\measnum$ acquired measurements into $\obsnum$ for map estimation and $\measnum-\obsnum$ for evaluating the quality of the estimate. To simplify the exposition,  the above notation assumes that the observed measurements are the first $\obsnum$ measurements ($\measind=1,\ldots,\obsnum$), but this is not necessary: the observed measurements can be any subset of the $\measnum$ measurements.
    \end{bullets}

    \subsubsection{Grid-aware RME}
    \label{sec:gridawarerme}

    \begin{bullets}
        \blt[intro]The second formulation adopts a \emph{grid discretization }approach, indirectly mitigating the effects of small-scale fading. This method is commonly employed by numerous radio map  estimators, including those based on deep learning, compressed sensing, and matrix completion.
        \blt[grid-aware]%
        \begin{bullets}%
            \blt[standard2grid]%
            \begin{bullets}%
                \blt[grid] Consider an $N_y \times N_x$ rectangular grid $\grid =
                    \left\{ \loc^{\grid}_{i, j}, i = 1, \ldots, N_y , j=1, \ldots, N_x
                    \right\}\subset \area$, where $\loc^{\grid}_{i,j} =[ \gridspacing (j-1),\gridspacing (N_y - i
                    )]\transpose$ and $\gridspacing$ is the grid spacing. This
                assignment facilitates identifying the grid with a matrix.
                \blt[assigment measurements to grid points]For each $\measind$, the
                measurement at $\loc_{\measind}$ is assigned to the nearest grid
                point. Subsequently, all measurements assigned to the $(i,j)$-th grid
                point are averaged to obtain $\powmeas{i,j}$. The resulting
                measurements are then arranged in the matrix $\powmeasmat \in \rfield^{N_y
                        \times N_x}$, whose $(i, j)$-th entry equals $\powmeas{i,j}$ if at
                least one measurement has been assigned to ${\loc^{\grid}_{i, j}}$ and
                an arbitrary value (e.g. 0) otherwise.
                \blt[mask]It is also convenient to
                form the \emph{mask} $\samplingmask \in \{0,1\}^{N_y \times N_x}$, which is an $N_y \times N_x$ matrix whose
                $(i, j)$-th entry equals 1 if at least one measurement has been
                assigned to ${\loc^{\grid}_{i, j}}$ and 0 otherwise.
                \blt[fading]Since $\powmeas{i,j}$ is the average of measurements
                acquired typically several wavelengths away, the contribution of $
                    \loss^{\text{SSF}} (\loc)$ is significantly
                reduced. The remaining terms in \eqref{received_power} evolve more slowly across space and, therefore, they do not vanish upon averaging.
            \end{bullets}

            \blt[formulation]
            \begin{bullets}%
                \blt[given]Given $\powmeasmat$ and $\samplingmask$,
                \blt[requested]the problem is to estimate the power map without the
                small-scale fading contribution
                for all $\loc \in \grid$. In the example of \eqref{received_power}, this can be expressed as
                \begin{align}
                    \label{eg:received_power}
                    \pownff \left( \loc \right) \define \txpower + \antgain - \loss^{\text{PL}} (\loc)  - \loss^{\text{S}} (\loc).
                \end{align}

            \end{bullets}
        \end{bullets}
    \end{bullets}
\end{bullets}

\subsection{Radio Map Estimators}
\label{sec:estimators}

The estimators considered in this work are outlined next. Implementation details are deferred to Sec.~\ref{sec:experiments}.

\subsubsection{$\neighbornum$-Nearest Neighbors ($\neighbornum$-NN)}
This algorithm simply averages the measurements collected at the
$\neighbornum$ locations with the smallest distance to the evaluation
point. Specifically, given $\loc$, let
$\neighborindfun_\neighborind(\loc)$ denote the index of the
$\neighborind$-th nearest point among
$\left\{\loc_{\measind}\right\}_{\measind=1}^{\obsnum}$. For example,
$\neighborindfun_1(\loc) = \argmin_{\measind}\| \loc_\measind - \loc
    \|$ whereas $\neighborindfun_{\obsnum}(\loc) = \argmax_{\measind}\|
    \loc_\measind - \loc \|$.  Although many variants exist, the simplest
is to obtain $
    \powest(\loc) = ({1}/{\neighbornum})\sum_{\neighborind=1}^{\neighbornum}\powmeas{\neighborindfun_{\neighborind}(\loc)}.
$

\subsubsection{Kriging} % \acom{combine this paragraph with the next ones}Kriging \cite{simpson2001kriging} involves creating a spatial model of radio signal strength using a set of measured signal strength values at specific locations. The model estimates signal strength at any location based on the signal strengths at the surrounding locations, considering the spatial correlation between these locations. The Kriging algorithm uses a covariance function to quantify the spatial correlation between signal strength values at different locations.
This \journal{ (see e.g.~\cite{beers2004kriging})} is a common spatial
interpolation technique extensively applied to
RME~\cite{alayafeki2008cartography,agarwal2018spectrum,shrestha2022surveying,boccolini2012wireless}. In
    {\em simple Kriging}, $\pow(\loc)$ is modeled for each $\loc$ as a
random variable whose spatial mean $\powmean(\loc) :=
    \expected[\pow(\loc)]$ and covariance $\cov[\pow(\loc),\pow(\loc')]$
are known for all $\loc$ and $\loc'$. In practice, these functions are
estimated from data.
\journal{Under model \eqref{eq:measmodel}, assume that $z_n$ is
    zero-mean with variance $\sigma_z^2$ and uncorrelated with $z_{n'}$
    for all $n'\neq n$ and with $\pow(\loc)$ for all $\loc$. Thus, the
    mean and covariance of the measurements are respectively
    $\expected[\powmeas{n}] = \powmean(\loc_n)$ and
    $\cov[\powmeas{n},\powmeas{n'}] = \cov[\pow(\loc_n),\pow(\loc_{n'})] +
        \sigma_z^2 \delta_{n,n'}$, where $\delta_{n,n'}$ equals $1$ if $n=n'$
    and $0$ otherwise. It can also be verified that
    $\cov[\pow(\loc),\powmeas{n}] = \cov[\pow(\loc),\pow(\loc_n)]$.} The
simple Kriging estimate is nothing but the linear minimum mean square
error (LMMSE) estimator of $\pow(\loc)$ based on \journal{the measurements}
$\powmeasvec\define[\powmeas{1},\ldots,\powmeas{\obsnum}]\transpose$:
\begin{align}
    \label{eq:lmmse}
    \powest(\loc) = \powmean(\loc) + \cov[\pow(\loc),\powmeasvec] {\cov}\inv[\powmeasvec,\powmeasvec](\powmeasvec - \expected[\powmeasvec]),
\end{align}
where \journal{$\cov[\powmeasvec,\powmeasvec]$ is the $\obsnum \times
        \obsnum$ matrix whose $(\measind,\measind')$-th entry is
    $\cov[\powmeas{\measind}, \powmeas{\measind'}]$ and
    $\cov[\pow(\loc),\powmeasvec]$ is the $1 \times \obsnum$ vector whose
    $\measind$-th entry equals $\cov[\pow(\loc),\powmeas{\measind}]$.}
\conference{$\cov[\powmeasvec,\powmeasvec]$ and
    $\cov[\pow(\loc),\powmeasvec]$ can be found by assuming that $z_n$ in
    \eqref{eq:measmodel} is zero-mean with variance $\sigma_z^2$ and
    uncorrelated with $z_{n'}$ for all $n'\neq n$ and with $\pow(\loc)$
    for all $\loc$ \cite{shrestha2022surveying}.}

\subsubsection{Kernel-based Learning}
Kernel-based estimators\journal{, grounded in the theory of reproducing-kernel
    Hilbert spaces~\cite{scholkopf2001}, are also common in RME; see
    \cite{romero2017spectrummaps} and references therein. They} return
estimates of the form $
    \powest(\loc)=\sum_{\measind=1}^{\obsnum} \krrcoef_\measind\kernelfun(\loc,\loc_\measind),
$
where the coefficients $\krrcoef_\measind$ depend on the specific
estimator and $\kernelfun$ is a \conference{ kernel function, such as
    a Gaussian radial basis function of width $\kernelwidth>0$, namely
    $\kernelfun(\loc,\loc') \define \exp\{ -
        \|\loc-\loc'\|^2/\kernelwidth \}$.}  \journal{user-selected symmetric
    positive (semi)definite function called kernel. A common example of
    kernel function is the so-called Gaussian radial basis function of
    width $\kernelwidth>0$, namely $\kernelfun(\loc,\loc') \define \exp\{
        - \|\loc-\loc'\|^2/\kernelwidth \}$.}  One of the simplest estimators
is \emph{kernel ridge regression} (KRR), where
$\{\krrcoef_\measind \}_{\measind=1}^{\obsnum}$ can be obtained in
closed form by solving
\begin{align}
    \minimize_{\{\krrcoef_\measind \}_{\measind=1}^{\obsnum}}
    \frac{1}{\obsnum}\sum_{\measind=1}^{\obsnum} \left|
    \powmeas{\measind} - \sum_{\measind'=1}^{\obsnum}
    \krrcoef_{\measind'}\kernelfun(\loc_\measind,\loc_{\measind'})\right|^2 +
    \regpar \sum_{\measind=1}^{\obsnum} \krrcoef_\measind^2,
    \label{eq:krrcoef}
\end{align}
where $\regpar>0$ is a regularization parameter.

% To reduce the influence of the kernel choice,
% \cite{bazerque2013basispursuit} proposed a \emph{multikernel}
% map estimator where the kernels in a given dictionary are combined
% based on the measurements.

\subsubsection{Deep Learning}
\label{sec:deeplearningestimators}
Numerous estimators leveraging CNNs have been
proposed  for RME~\cite{parera2020tiltdependent, imai2019radiopredictioncnn, iwasaki2020transferbasedpower, teganya2020rme, krijestorac2020deeplearning, thrane2020model, niu2018recnet, ratnam2020fadenet, levie2019radiounet, han2020power, shrestha2021deep,
    shrestha2022surveying}. A typical approach is to concatenate
$\powmeasmat$ and $\samplingmask$ to form a $2 \times N_y \times N_x$
tensor that is passed as an input to a neural network, which returns an
$N_y \times N_x$ matrix $\powestmat$ whose $(i,j)$-th entry is an
estimate for $\pownff(\loc^{\grid}_{i,j})$. To simplify subsequent
expressions, this estimate will still be denoted as
$\powest(\loc^{\grid}_{i,j})$.

\section{Data Collection and Preparation}
\label{sec:collection}

\begin{bullets}%
    \blt[estimation instance] To train and test the radio map estimators, it is necessary to generate  a large number of \emph{estimation instances}, which is the term that will be used here to refer to a set of $\measnum$ measurements in an area $\area$.
    For specificity, $\area$ is assumed to be a rectangle  $\area$ of fixed size  $\areasidey \times \areasidex$.
    \blt[naive approach]A naive approach would be that each estimation instance comprises $\measnum$ measurements collected in a different $\areasidey\times \areasidex$ geographical region. However, this would necessitate   a very large number of measurements.
    \blt[large map approach]
    \begin{bullets}%
        \blt[patches]Instead, a more efficient approach is to collect a large number
        $\lgmeasnum$
        of measurements in a large rectangle $\lgarea=[0,\lgareasidey]\times[0,\lgareasidex]$, where $\lgareasidey\gg \areasidey$ and $\lgareasidex\gg \areasidex$. The set of these $\lgmeasnum$ measurements will be referred to as a \emph{measurement set}; cf. Table~\ref{tab:terms}.
        Then, a large number of estimation instances can be generated by forming $\areasidey\times \areasidex$ rectangular patches inside $\lgarea$ and selecting the measurements inside each patch.
        \ifroom{Observe that, in this approach, each of the $\lgmeasnum$ measurements is generally used in multiple estimation instances.}

        \blt[grid] As will become clear, to study grid-aware estimators and the impact of small-scale fading it is useful that the  measurements in a patch are aligned to some extent with the grid; cf. Sec.~\ref{sec:gridawarerme}. To accommodate this need, an $\lggridy\times\lggridx$ grid with spacing $\gridspacing$ is defined on $\lgarea$, meaning that $\lgareasidex=\lggridx \gridspacing$ and $\lgareasidey=\lggridy \gridspacing$. To ensure that the measurements in each patch remain aligned with the grid,
        the patches are formed by randomly drawing the location of the bottom-left corner such that its $x$ and $y$-coordinates are integer multiples of $\gridspacing$. This yields at most   $(\lggridx-\gridx+1)(\lggridy-\gridy+1)$ distinct patches.

    \end{bullets}%

    \blt[overview]The rest of the section details the procedures to collect the measurement sets, which are grouped into two datasets.

\end{bullets}%

\subsection{USRP Dataset}

\begin{figure}[t]
    \centering
    \includegraphics[width=2.5in]{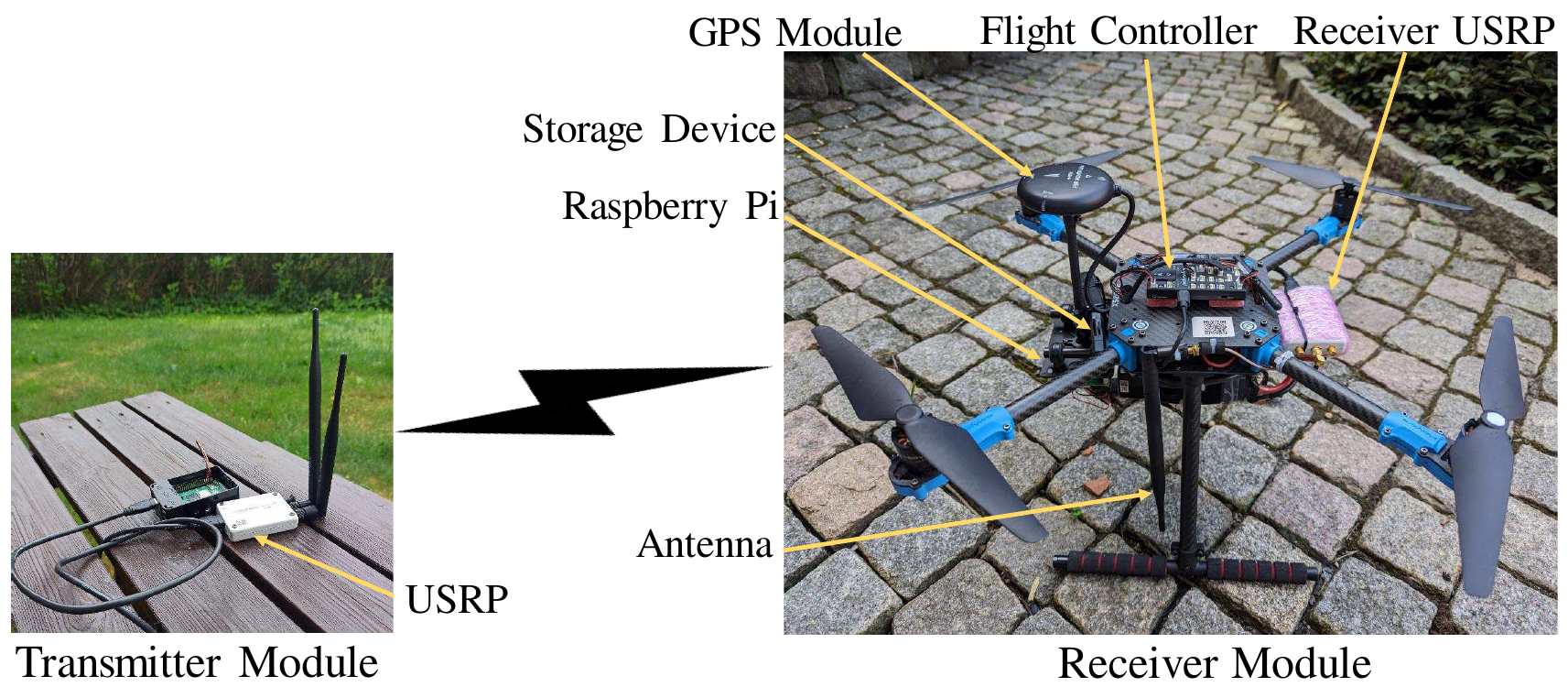}
    \caption{
        Transmit and receive  modules.}
    \label{fig:system}
\end{figure}

The first dataset comprises around 320,000 measurement locations arranged into 26 measurement sets. To be able to collect such a large amount of measurements in a reasonable time, a custom data collection system had to be developed.
%  and was collected using a transmitter stationed at a fixed ground location and a receiver  mounted on a UAV.

\subsubsection{Transmitter}

\begin{bullets}
    \blt[USRP]The transmitter (see Fig.~\ref{fig:system}) is a Universal
    Software Radio Peripheral (USRP) B205 mini-i \journal{operated by a
        Raspberry Pi, both of which are connected through a battery
        elimination circuit (BEC) to a battery so that they can be deployed on
        the field.  The antenna is a 21 cm-long monopole with an operating
        range of 824-960 MHz and 1710-1990 MHz. }\conference{with a monopole
        antenna of 21 cm.}
    \blt[transmission signal/parameters] The USRP transmits an \emph{orthogonal
        frequency-division multiplexing} (OFDM) signal with
    %~\cite{jiang2008overview} with
    %
    \conference{1024 subcarriers, out of which the central 600 are
        used. Since the sampling rate is 5 MHz, the effective bandwidth
        becomes roughly 2.93 MHz. A frame that contains 12 OFDM symbols is
        repeatedly transmitted with a carrier frequency of 918 MHz. }%
    \begin{bullets}%
        \blt[FFT Len/subcarriers]1024 subcarriers,
        \blt[ch. BW] and sampling rate of 5 MHz,
        \blt[subcarrier spacing] which yields a subcarrier spacing of 15 KHz.
        \blt[occupied subcarriers] The 600 central subcarriers are
        modulated with QPSK symbols that encode a pseudo-random bit
        sequence. The remaining subcarriers are unused, which yields an
        effective bandwidth of roughly $600/1024 \cdot 5$ Mhz $\approx 2.93$ MHz.
        \blt[cyclic prefix] The cyclic prefix is of length 256.
        \blt[frame] A frame that contains 12 OFDM symbols is repeatedly transmitted  with a
        \blt[Center freq] carrier frequency of 918 MHz. This band is mostly unused in Norway.
        \ifroom{\blt[power] The power amplifier (PA) was adjusted for maximum
            transmission power of  -20 dBm, but the actual transmit power is
            considerably lower since the transmitted signal is scaled so that only one
            in 10,000 measurements saturates. This keeps the PA in its linear
            region and avoids problems associated with the peak-to-average
            power ratio (PAPR) of OFDM signals.}
    \end{bullets}

\end{bullets}

\subsubsection{Receiver}

%To minimize the effect of self-emissions of the UAV~\cite{pienaar2016rfshielding}, the receiver antenna is positioned in the middle of the frame facing downward.

\begin{bullets}
    \blt[UAV]
    \begin{bullets}
        \blt[vehicle] The receiver, which is a USRP B205 mini-i with the same
        antenna as the transmitter (Fig.~\ref{fig:system})\ifroom{ placed away from the motors to minimize the impact of self-emission \cite{pienaar2016rfshielding}}, is installed on a
        quadcopter with a Raspberry Pi companion computer\ifroom{ and an
            external SD card for data storage}. The quadcopter was assembled using
        a Holybro X500 v2 frame and a Pixhawk 4 flight controller (FC) that
        runs a PX4 autopilot.
        The autopilot estimates the vehicle location  by
        fusing the measurements of an \emph{inertial measurement unit} (IMU) and  an RTK module, resulting in an
        accuracy of around 30 cm.
        \ifroom{The FC communicates with a ground control
            station (GCS) through a telemetry link at 433 MHz.}
        %

        % \blt[Localization] The GCS is connected to a real-time kinematic (RTK)
        % base station H-RTK F9P and sends phase corrections to the FC through
        % the telemetry link so that the global navigation satellite system
        % (GNSS) position estimates of the autopilot have an accuracy of around
        % 30 cm. The autopilot fuses these estimates with the measurements of an
        % IMU to obtain accurate location estimates at a high
        % rate. The companion computer then retrieves these estimates from the FC
        % through a USB connection. %
        % %

        \begin{figure}[t]
            \centering
            \includegraphics[width=3.5in]{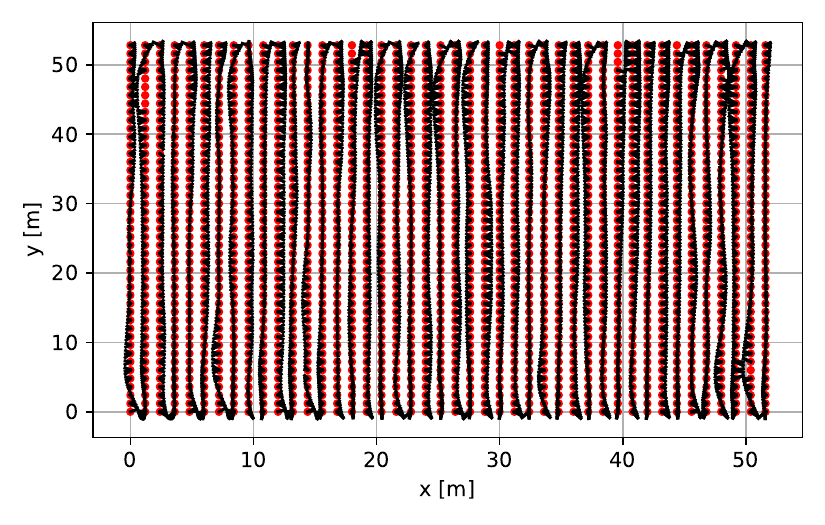}
            \caption{Grid quantization (cf. Sec.~\ref{sec:gridawarerme}) of the measurement
                locations in one of the collected measurement sets in the USRP dataset. Red dots denote
                grid points and small black lines connect each measurement location to
                its nearest grid point.}
            \label{fig:grid_quantization}
        \end{figure}

    \end{bullets}

    \blt[comms. part]

    \begin{bullets}

        \blt[acquisition] Every second, the receiver module captures approximately 15 uniformly-spaced blocks of 100,000 samples.
        \ifroom{ a rate that is limited by the writing speed
            of the storage SD card.}Given that the speed is 4 m/s, the average
        distance between sample blocks is around 27 cm, slightly smaller than
        the wavelength of the transmission. This is useful to
        average out small-scale fading, as described in Sec.~\ref{sec:problem}.
        Together with each sample block, the receiver stores the corresponding location
        estimate.

        \blt[signal power estimates] The  sample blocks are processed offline with an  algorithm designed for this project to coherently estimate the received signal power, which requires accurate synchronization but minimizes measurement error due to noise and interference.
        \ifroom{The estimates and compensates for the carrier frequency offset and frame offset,
            removes the cyclic prefixes, correlates the transmitted QPSK symbols
            with the received sequence in the frequency domain to obtain an estimate
            per OFDM symbol of the channel scaled by the transmitted power,
            estimates the phase offset, and combines the aforementioned estimates
            into a single one. The power is estimated by adding the square
            magnitudes of used subcarrier observations and yields what is referred
            to as a power measurement $\powmeas{\measind}$.}

    \end{bullets}

\end{bullets}

\subsubsection{Data Collection Procedure and Postprocessing}
\label{sec:usrpdsdc}

\begin{figure}[t]
    \centering
    \includegraphics[width=\linewidth,height=4.85cm]{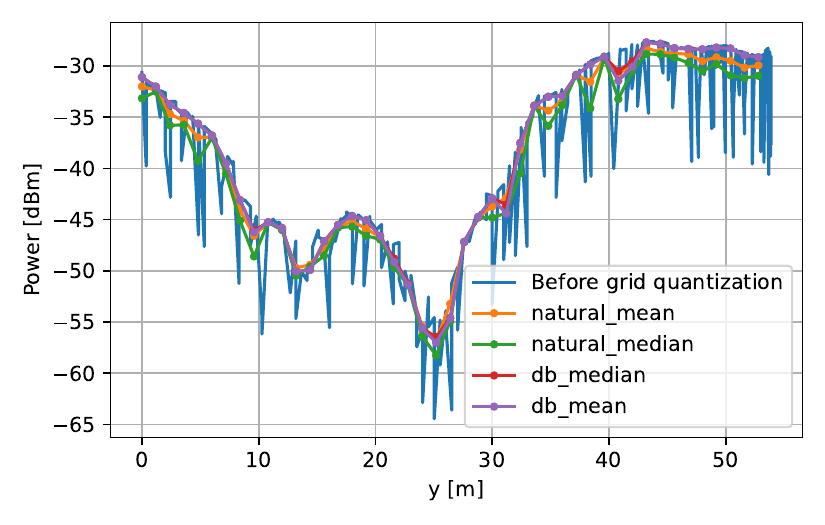}
    \caption{To analyze approaches for averaging out small-scale fading, this figure represents measurements in the 41-st column of
    Fig.~\ref{fig:grid_quantization} vs. the $y$-coordinate of their
    location.  Orange, Green, Red, Purple: Measurements $\{\powmeas{i,j}\}_{j}$ for a fixed
    $j=41$ vs. the $y$-coordinate of $\loc^\grid_{i,j}$ with grid quantization modes natural mean, dB mean, natural median, and dB median respectively. Blue:
    measurements $\{\powmeas{\measind}\}_{\measind}$ assigned to the
    points in $\{\loc^\grid_{i,j}\}_{j}$.  }
    \label{fig:fading}
\end{figure}

\begin{bullets}

    \blt[acquisition]
    \begin{bullets}
        \blt[environment]For safety reasons,  data was collected in a flat agricultural terrain away from residential areas.
        \blt[datasets] To account for different propagation conditions, two kinds of
        measurement sets were collected. The first kind comprises 8
        \emph{line-of-sight} (LOS) measurement sets. For each one,  the
        transmitter was placed at a different distance from the center of the
        measurement region. The second kind comprises 18 \emph{intense shadowing}
        measurement sets, where abrupt shadowing patterns were created by
        arranging metallic reflectors around the transmitter in a different configuration for each measurement set.
        In both kinds,  $\lgmeasnum$ is approximately 12,000 and the transmitter was placed about 50 cm above the
        ground.

        \blt[trajectory]%
        \begin{bullets}%
            \blt[height]To facilitate experiments with
            grid quantization, the UAV follows a trajectory with a height of 7 m
            \blt[spacing] that comprises parallel lines spaced by $\gridspacing$. Fig.~\ref{fig:grid_quantization} illustrates this trajectory for one of the measurement sets after removing the turnarounds.
            \blt[yaw] To minimize changes in the channel due to the UAV frame,
            the yaw angle is kept constant.
            \blt[explain why region is square?]To maximize the number of patches, the region was selected such that $\lggridx=\lggridy$; cf. the beginning of Sec.~\ref{sec:collection}.
            \blt[explain why we chose 1.2 m spacing?]The size of the region is limited by the battery life of the UAV. With the adopted configuration, the UAV can fly for about 10 minutes. With a speed of 4 m/s, this means that a distance of 2400 m can be covered. If one wishes to allow for the creation of a grid with $\lggridx=\lggridy$ and spacing $\gridspacing$, the UAV must travel a distance of $(\lggridx^2-1) \gridspacing$. Therefore, $\gridspacing$ and $\lggridx$ must be selected such that $(\lggridx^2-1) \gridspacing = 2400$. Given that $\lgareasidex=\lggridx \gridspacing$, there is a trade-off between the number of patches (determined by $\lggridx$) and the size of the region. To attain a reasonable value for both quantities, $\lggridx$ was set to 45, which yields $\gridspacing \approx 1.2$ m.
        \end{bullets}

    \end{bullets}

    \blt[processing]

    \begin{bullets}

        \blt[cleaned] Each measurement  set is then cleaned to keep only a
        subset of the measurements that lie inside a rectangular region, which
        removes the turnaround parts of the trajectory as well as the path
        from the take-off location to the first waypoint and from the last
        waypoint to the landing location.

        \blt[grid alignment] The geodetic coordinates stored by the receiver
        are then converted to Cartesian coordinates and the latter are
        translated and rotated so that the measurement locations align as much
        as possible with the grid. This means that most grid points should be assigned to at
        least one measurement location. However, due to the effects of wind or
        the UAV momentum when changing direction, some of the grid points may
        not be assigned to any measurement;
        cf. Fig.~\ref{fig:grid_quantization}.  %
        \conference{The measurement
            locations are then rotated and translated to maximize alignment with
            the grid; cf. Fig.~\ref{fig:grid_quantization}. Wind sometimes results
            in some grid points without assigned measurements.}
        \blt[figure]Yet, the alignment is sufficiently good to allow the
        visualization of each measurement set as in Fig.~\ref{fig:3D_true_map},
        where just the grid quantization procedure from Sec.~\ref{sec:problem}
        was applied. Observe that the ripple effect of small-scale fading (cf. $
            \loss^{\text{SSF}} (\loc)$ in \eqref{received_power}) has not been
        totally suppressed  as a result of  grid quantization.

        \blt[averaging function]To entirely
        mitigate this effect, one could adopt a larger grid spacing $\gridspacing$,
        but this would come at the cost of reduced
        spatial resolution. Another possibility is to consider alternative functions for  combining the measurements assigned to each grid point; cf. Sec.~\ref{sec:gridawarerme}. To this end,  Fig.~\ref{fig:fading}, illustrates the
        measurements assigned to the grid points in  a column of Fig.~\ref{fig:grid_quantization} before
        and after grid quantization using various combination functions. Specifically, $\powmeas{i,j}$ is
        obtained by taking the mean or median of the  measurements assigned to
        $\loc^{\grid}_{i, j}$ either in natural or dB units. \emph{Averaging the measurements in dB units is seen to result in  the best averaging of small-scale fading} and, therefore, it will be the approach  adopted in the sequel. Yet,  its effectiveness is limited, which calls for further research to devise more effective combining functions.

    \end{bullets}

\end{bullets}

% [small-scale fading mitigation]
% To mitigate the effects of small-scale fading, the conventional
% approach often involves conducting measurements at multiple locations,
% which can be costly or time-consuming.
% %
% Instead, this work adopts averaging across the measurements in the
% frequency domain that are collected at the same location, which is
% relatively equivalent to averaging across the spatial domain with
% the same frequency.

\subsection{4G Dataset}
The second dataset is intended to assess RME in cellular setups. It comprises a large number of measurement sets, each one for a different metric and 4G cell in two geographical areas.
\begin{bullets}
    \blt[Motivation 4G:]The reason for focusing on 4G systems  is their wider coverage relative to their 5G counterparts, especially away from urban areas, where UAVs are allowed to fly.
\end{bullets}

\subsubsection{Transmitter}
The transmitters are the  base stations deployed by a cellular operator in a real-world 4G network. \begin{changes}
    % Conventional sector antennas are used.
\end{changes}

\subsubsection{Receiver}
\begin{bullets}
    \blt[UAV]
    \begin{bullets}
        \blt[vehicle] The receiver is a Quectel RM510Q-GL modem on board a DJI Matrice 300 RTK UAV. The localization module is the  built-in RTK rover module of the UAV, which ensures highly accurate geo-referencing of the  LTE data.

    \end{bullets}
    \blt[communications Part]
    \begin{bullets}
        \blt[acquisition] The   magnitudes  measured at each location comprise the \emph{reference signal received power} (RSRP), the \emph{received signal strength indicator} (RSSI), and the
        \emph{reference signal received quality} (RSRQ) for the serving cell as well as for 8 neighboring cells. These metrics are measured  and stored on board together with the  location estimates.
    \end{bullets}
\end{bullets}

\begin{figure}[t]
    \centering
    \includegraphics[width=3.5in]{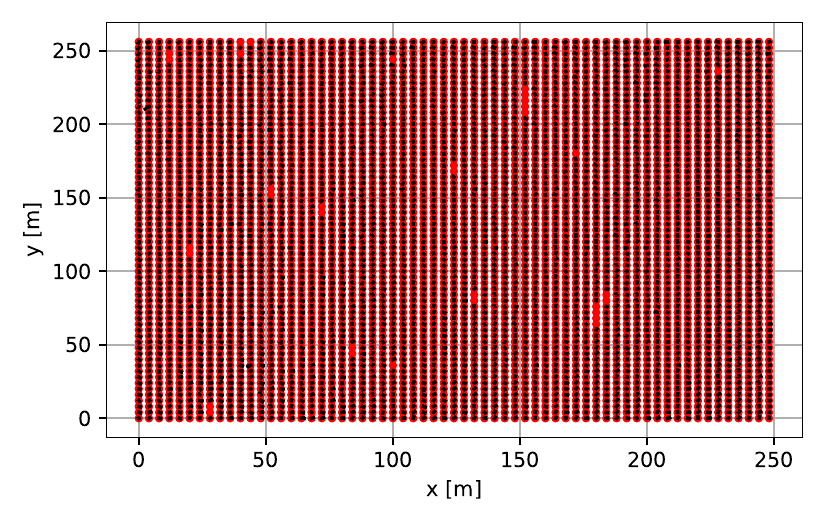}
    \caption{Grid quantization of the measurement
        locations in one of the collected measurement sets of the 4G dataset. Red dots denote
        grid points and small black lines connect each measurement location to
        its nearest grid point.}
    \label{fig:grid_quantization_mabia}
\end{figure}

\subsubsection{Data Collection Procedure and Postprocessing}
\begin{bullets}
    \blt[acquisition]
    \begin{bullets}
        \blt[environment]The data is collected in two remote areas away from agricultural and residential facilities.
        \blt[regions]In both cases, $\lgareasidex=252$ m and $\lgareasidey=260$ m and
        \blt[Trajectory]%
        \begin{bullets}%
            \blt[height] the height is 20 m.
            \blt [grid] A rectangular grid with $\gridspacing=4$ m is constructed and the UAV   hovers at each grid point to collect 5 measurements per metric.
        \end{bullets}%
        \blt[datasets]93 cells are measured at least at one point for the first area and 31 for the second one.
    \end{bullets}%
    \blt[postprocessing]%
    \begin{bullets}%
        \blt[how rsrp metric obtained]The measurement locations are  rotated and translated to maximize the alignment with the grid;
        \blt[figure of grid aligment]cf. Fig.~\ref{fig:grid_quantization_mabia}.
    \end{bullets}
\end{bullets}

\section{Experimental Study}
\label{sec:experiments}

\newcommand{\mcfigure}[3]{
    \begin{figure}[t]
        \centering
        \includegraphics[width=\linewidth]{figs/MC_exp/experiment_48091-#1.pdf}
        \caption{#3}
        \label{#2}
    \end{figure}
}

\begin{bullets}
    \blt[overview]This section describes the experiments carried out with the collected data. Table~\ref{tab:terms} summarizes the main terminology.
    \blt[data]%
    \begin{bullets}%

        \blt[estimation instances] For training and evaluating the performance metrics via  Monte Carlo (MC)
        averaging, each estimation instance is generated by selecting the measurements inside a ``patch'' of the region $\lgarea$ corresponding to a measurement set selected uniformly at random (cf. Sec.~\ref{sec:collection}).
        With
        $\left\{(\loc_{\measind},
            \powmeas{\measind})\right\}_{\measind=1}^{\measnum}$ denoting these measurements, 
        $\loc_{\measind}$, $\powmeas{\measind}$, and $\measnum$ can be
        thought of as random variables that take new values for each patch or
        MC iteration. For simplicity, patches are square, i.e. $\areasidex=\areasidey\define\areaside$.

    \end{bullets}

    \begin{figure*}[t!]
        \centering
        \includegraphics[width=1.05\linewidth]{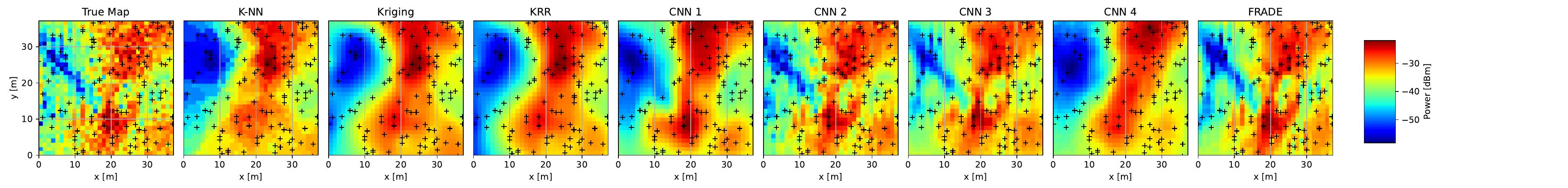}
        \caption{True map and  map estimates produced by the considered
            map estimators for an estimation instance drawn from the USRP dataset when $\areaside=38.4$~m, $\gridspacing = 1.2$, and $\obsnum = 120$. Black crosses denote measurement locations.}
        \label{fig:estimation_example}
    \end{figure*}

    \blt[estimation example]Fig.~\ref{fig:estimation_example} depicts 8 map estimates for a randomly selected patch of
     1,475 m$^2$ using the  USRP data set; see Sec.~\ref{sec:usrpexperiments}. The number of observations $\obsnum= 120$ was selected so that all estimates are of a reasonable
    quality.

    % \blt[remaining] The rest of this section considers several setups to
    % assess the performance of the considered estimators in both of the
    % problem formulations in Sec.~\ref{sec:problem}.

\end{bullets}

\subsection{Compared Algorithms}
The compared algorithms include three \emph{traditional} (non-CNN) estimators and 4 state-of-the-art CNN-based estimators.

\subsubsection{Non-CNN Estimators}
\label{sec:sim:nonCNNestimators}
This includes  K-NN, simple Kriging, and  KRR, which are general-purpose function regression algorithms that have been applied to RME; cf. Sec.~\ref{sec:estimators}.
\begin{bullets}
    \blt[common training]The parameters of these  estimators are obtained via exhaustive search over a grid of candidate values by minimizing a sample estimate of the RMSE in~\eqref{eq:rmsestandarduniform} on the training estimation  instances of the corresponding experiment.  The range of  possible values was set to  include the optimal value in its interior. This ensures an \emph{fair comparison} with CNN estimators.  The specifics of each algorithm are detailed next.

    \blt[Parameters used for training]
    \begin{itemize}
        \item\cmt{KNN}K-NN is trained for
        $\neighbornum \in [2, 3,\ldots, 13]$.
        \item\cmt{Kriging}Simple Kriging adopts the correlated shadowing model from~\cite{gudmundson1991correlation}. This model can be defined     by
        $\expected[\pow(\loc)]=0$ and
        $\text{Cov}(\loss^{\text{S}} (\loc), \loss^{\text{S}} (\loc\prime)) =
            \ushadvar2^{-\|\loc - \loc^{\prime}\|/\dist}$, where
        $\ushadvar$ captures variability and  $\dist$ is the distance at which the correlation decays
        to $1/2$; see~\cite{shrestha2022surveying} for details. The values used for training are
        $\ushadvar \in [0.01^2, 0.11^2, \ldots,0.91^2]$  and $\dist \in [50, 100, \ldots, 600]$~m. 
    The term $ \loss^{\text{SSF}} $ is assumed spatially white and the rest of terms deterministic unknown.
        \item\cmt{KRR} KRR is trained for all combinations of   $\regpar \in [10^{-12}, 10^{-11}, \ldots, 10^{-1}]$ and kernels in a set that  includes both Gaussian and Laplacian kernels of width  $\kernelwidth \in [20, 30, \ldots, 150]$.
    \end{itemize}
\end{bullets}

\subsubsection{CNN Estimators}
Four CNN estimators from the literature were considered.
\begin{bullets}%
    \blt[common]They construct a grid with
    $N_x=N_y=\areaside/\gridspacing$  and obtain
    $\powmeasmat$ and $\samplingmask$ as described in
    Sec.~\ref{sec:gridawarerme}. A forward pass yields $\powestmat$, which
    provides estimates of the map at the grid locations. Estimates  off the grid points are obtained as the entry of $\powestmat$ corresponding to the nearest grid point.
    \blt[training data]To generate training input-output pairs,
    $\powmeasmat$ and $\samplingmask$ are first obtained from all the $\measnum$ measurements in a patch. The input is then obtained by setting a randomly selected subset of entries of both $\powmeasmat$ and $\samplingmask$ equal to 0, leaving just $\obsnum$ non-zero entries; see details in
    Sec.~\ref{sec:gridawaremetrics}. The targets are the complete  $\powmeasmat$.

    %
    % \blt[Data augmentation]To artificially expand the size of the training dataset, data augmentation is employed by randomly applying operations such as horizontal flipping, vertical flipping, and a 90-degree rotation to the training patches.
    %
    \blt[CNN parameters]The CNNs were implemented in TensorFlow and
    trained with the Adam optimizer with a constant learning
    rate of $10^{-4}$ and batch size of 200. The   CNN-based estimators mainly differ in their network architectures.

    \blt[specifics to each network]
    \begin{itemize}
        \item\cmt{CNN-1}CNN 1 is the network in~\cite{shrestha2022surveying}, which has an autoencoder architecture with
        % 26 layers including
        % convolutional, transpose convolutional, max pooling, and upsampling
        % layers, all with  \textit{leaky relu} activations. With 
        60 M trainable parameters. This is the network with  the highest
        complexity  here.
        \item\cmt{CNN-2}CNN 2 is the  U-Net from~\cite{levie2019radiounet}. Four
        layers were removed, as was required to accommodate the grid size used here. With 9 M
        trainable parameters in total, it features the second highest
        complexity.
        \item\cmt{CNN-3}CNN 3 follows the U-Net
        architecture in~\cite{krijestorac2020deeplearning} \journal{with 20 layers but} with
        \textit{leaky ReLU} activations, since the original tanh activation resulted in poor performance. With 4.6 M trainable parameters, it has the third highest complexity.
        \item\cmt{CNN-4}CNN 4 is the completion autoencoder from~\cite{teganya2020rme}. With 142 K trainable parameters, it has the lowest complexity.

    \end{itemize}

    % All estimators were trained separately for each value of  $\areaside$, $\obsnum$ and dataset used in the experiments.

\end{bullets}

\subsubsection{Combination of non-CNN and CNN Estimators}
The upcoming experiments suggest the convenience of combining both classes of estimators. The resulting family of \emph{hybrid estimators} will be discussed in Sec.~\ref{sec:hybrid}.

\subsection{Performance Metrics}
\label{sec:metrics}

To assess performance  in a variety of scenarios
that reflect the relevant aspects of the main applications of RME, this section
introduces four different  metrics  and three spatial sampling approaches to select the observations among the available measurements.

\subsubsection{Grid-agnostic Estimation}
\begin{bullets}
    \blt[overview]When solving the problem in
    Sec.~\ref{sec:gridagnosticrme}, it is natural to quantify performance
    by the estimation error at the locations of unobserved measurements, not necessarily on a grid.
    \blt[performance evaluation]
    \begin{bullets}
        \blt[observations] Specifically, at each MC iteration, a patch is randomly generated (see Sec.~\ref{sec:collection}) and the set of measurements
        $\measset\define\left\{(\loc_{\measind},
            \powmeas{\measind})\right\}_{\measind=1}^{\measnum}$ in the patch is split
        into two subsets by partitioning the index set
        $\measindset\define \{1,2,\ldots,\measnum\}$ into $\obsmeasindset$ and
        $\nobsmeasindset$, that is,
        $\obsmeasindset\cup\nobsmeasindset=\measindset$ and
        $\obsmeasindset\cap\nobsmeasindset=\emptyset$. The cardinality
        $\obsnum\define |\obsmeasindset|$ is fixed.
        \blt[estimates] The measurements with index in $\obsmeasindset$ are
        passed to each estimator and the returned map estimate
        $\powest(\loc)$ is evaluated at the locations
        $\{\loc_\measind\}_{\measind \in \nobsmeasindset}$.  The RMSE can then
        be defined as \blt[metric: standard-uniform]
        \begin{align}
            \label{eq:rmsestandarduniform}
            \text{RMSE} \define \sqrt{
                \frac{1}{|\nobsmeasindset|}\expected\left[ \sum_{\measind\in\nobsmeasindset}|
                    \powmeas{\measind}-\powest(\loc_\measind)
                    |^2\right]
            },
        \end{align}
        where the expectation $\expected$ is over patches and index
        sets $\obsmeasindset$ sampled uniformly at random without replacement
        from $\measindset$.

        \blt[metric: standard-grid]Besides  \eqref{eq:rmsestandarduniform}, another metric will be considered to capture the clustered nature of  measurement locations in certain applications. This situation arises e.g. in  cellular networks in which numerous measurements are collected, but only where the users are located. To capture this effect, define the
        metric $\text{RMSE}_\text{G}$ as in \eqref{eq:rmsestandarduniform} but
        with a different distribution for $\obsmeasindset$: first  create a
        rectangular grid with spacing $\gridspacing$ and assign  each measurement location  to the nearest grid point. Then select $\obsnum$ grid points uniformly at random and form $\obsmeasindset$  by
        collecting the indices of the measurements assigned to the selected
        grid points. Note that, in this case, $\obsnum$ no longer indicates the number of observed measurements, but the number of \emph{measurement clusters}.

    \end{bullets}

\end{bullets}

% \mcfigure{0}{fig:agnostic16}{Grid-agnostic performance metrics for the
%     compared estimators vs. the number of observations when
%     $\areaside=19.2$ m. The RMSE$_\text{G}$ of the multikernel estimator exceeds
%     10 dB and therefore it is not visible in the bottom figure. }

% \mcfigure{1}{fig:aware16}{Grid-aware performance metrics for the compared
%     estimators vs. the number of observations when  $\areaside=19.2$~m.}

% \begin{figure}[t]
%     \centering
%     \includegraphics[width=\linewidth]{figs/MC_exp/experiment_48102.pdf}
%     \caption{Two of the considered performance metrics vs. the number of observations when  $\areaside=38.4$~m. }
%     \label{fig:both32}
% \end{figure}

\subsubsection{Grid-aware Estimation}
\label{sec:gridawaremetrics}
\begin{bullets}
    \blt[overview] When solving the problem in
    Sec.~\ref{sec:gridawarerme}, it is natural to evaluate the performance
    on $\grid$.
    \blt[performance evaluation]%
    \begin{bullets}%
        \blt[observations] To this end, at each MC iteration, a patch is drawn at random and $\powmeasmat$
        and $\samplingmask$ are constructed as in Sec.~\ref{sec:gridawarerme}. Let
        $\measindset\subset\{1,\ldots,N_y\} \times \{1,\ldots,N_x\}$ denote
        the set of pairs $(i,j)$ such that $[\samplingmask]_{i,j}=1$. As
        before, $\measindset$ is partitioned into $\obsmeasindset$ and
        $\nobsmeasindset$, where $\obsnum\define |\obsmeasindset|$ is given.
        % , but it should be clear that, in Sec.~\ref{sec:gridagnosticrme}, it refers to the number of observed measurements, whereas here it refers to the number of observed grid points. The meaning at each case can be inferred from the considered metric.
        % 
        \blt[estimates] Each estimator receives $\obsmeasindset$ and
        $\{\powmeas{i,j}\}_{(i,j)\in\obsmeasindset}$ and produces an estimate
        $\powestmat$.
        % of $\pownff$ at points
        % $\{\loc^{\grid}_{i,j}\}_{(i,j)\in\nobsmeasindset}$. 
        \blt[metric: grid nobs]To quantify performance,  consider
        \begin{align}
            \label{eq:rmsenobs}
            \text{RMSE}_{\gridnobs} \define \sqrt{
                \frac{1}{|\nobsmeasindset|}\expected\left[\sum_{(i,j)\in\nobsmeasindset}|
                    \powmeas{i,j}-\powest(\loc^{\grid}_{i,j})
                    |^2\right]
            },
        \end{align}
        where  $\expectation$ is over patches and over $\obsmeasindset$,
        which is drawn uniformly at random over $\measindset$ without replacement.

        \blt[metric: grid all] Various estimators may place different emphasis on the spatial smoothness of their estimates. Thus, it is also
        insightful to consider $\text{RMSE}_{\gridall}$, where $\obsmeasindset$ is
        drawn in the same way as in $\text{RMSE}_{\gridnobs}$ but the
        evaluation takes place at all grid points in $\measindset$, that is,
        $\nobsmeasindset$ in \eqref{eq:rmsenobs} is replaced with
        $\measindset$. Note that $\text{RMSE}_{\gridall}$ is the typical metric used for training CNN-based estimators in the RME literature.

    \end{bullets}

    % \begin{figure}[t]
    % \centering
    % \includegraphics[width=\linewidth]{figs/rmse_vs_num_meas_grid_all.pdf}
    % \caption{.}
    % \label{fig:gridrmse_vs_nummeas}
    % \end{figure}

\end{bullets}

\subsection{Estimation Performance Analysis}
\label{sec:performance}

This section tests a broad range of estimators on the USRP and 4G datasets. MC experiments are used to obtain the performance metrics introduced in Sec.~\ref{sec:metrics}.

\subsubsection{Experiments with USRP Data}
\label{sec:usrpexperiments}
\begin{bullets}
    \blt[overview]This section considers
    \blt[training/test]the 18 intense shadowing measurement sets in the USRP dataset. 16 of them  are used for training and the remaining 2 for testing.
    \begin{bullets}%
        \blt[training]%
        \begin{bullets}%
            \blt[Number of unique maps]A total of 40,000 estimation instances are formed by first picking a training measurement set uniformly at random and then randomly choosing a patch as described earlier.
            \blt[number of blocks/per patch]Each of these estimation instances is partitioned into an observed and an unobserved set  5 times as described in Sec.~\ref{sec:gridawaremetrics},
            \blt[Total training Examples]resulting in a total of 200,000 training examples.
        \end{bullets}
    \end{bullets}%

    %
    %\blt[patch size] The
    \blt[figures-USRP]
    \begin{bullets}
        \blt[16 x 16]
        \begin{bullets}
            \blt[10 to 100 meas]
            \begin{bullets}
                \blt[description]Figs.~\ref{fig:usrp_agnostic16_10_to_100} and
                \ref{fig:usrp_aware16_10_to_100} depict all four performance
                metrics when the patch side is $\areaside=19.2$ m, which results
                in  an area of 369 m$^2$ and $\gridx=\gridy=\areaside/\gridspacing=16$. This value of $\areaside$ ensures that
                 the training set is sufficiently rich for CNN
                training. To generate each training example, $\obsnum$ was drawn
                uniformly at random between 10 and 100. A curve labeled as \emph{FRADE} is  included in each of these and other upcoming figures but they will not be discussed before Sec.~\ref{sec:hybrid}. This is  because it corresponds to an algorithm that has not been considered in the literature.

                The four x-axes in Figs.~\ref{fig:usrp_agnostic16_10_to_100} and
                \ref{fig:usrp_aware16_10_to_100} are equivalent representations of the  same values of $\obsnum$. 
                In the top subfigure of Fig. \ref{fig:usrp_aware16_10_to_100},
                $\obsnum$ is normalized to reflect the fraction of observed grid points.
                In the top subfigure of Fig.~\ref{fig:usrp_agnostic16_10_to_100}, $\obsnum$ is scaled by $1/(\areaside^2/\lambda^2)=\lambda^2/\areaside^2$ to yield  the \emph{normalized measurement density}, that is, the number of observed measurements per area of $\lambda^2$ m$^2$. This is because, in homogeneous media, electromagnetic fields depend  on distances only through the ratio between these distances  and the wavelength. Thus, this normalization makes the results approximately independent of the carrier frequency.\footnote{The results are not totally independent of the carrier frequency because of the  localization error and the frequency dependence of the electromagnetic properties of the  environment.}

                \blt[observations]
                \begin{bullets}
                    \blt[overall error]Overall, the error of most estimators for a sufficiently large $\obsnum$ is
                    around 5 dB for the grid-agnostic metrics and 3 dB for the grid-aware metrics. This agrees with intuition: grid metrics rely on grid quantization, which \emph{averages out
                        small-scale fading}. This reduces the spatial variability of the target $\pow(\loc)$, which
                    renders it an easier function to estimate.
                    \blt[grid-agnostic metrics]Among grid-agnostic metrics,  it is observed in Fig.~\ref{fig:usrp_agnostic16_10_to_100} that
                    RMSE$_\text{G}$ is generally smaller than RMSE. The reason is that the measurement locations are more clustered for  RMSE$_\text{G}$. This  encourages estimators to fit a local average of the measurements in each cluster, which  implicitly  averages out measurement noise and small-scale fading,  leading to spatially smoother estimates.
                    The best estimators approximately attain their minimum RMSE$_\text{G}$
                    with 60 observations, which amounts to one observation for every
                    6.1 $\text{m}^2$,  or equivalently, 17 observations for each area unit of 1000 $\lambda^2$ m$^2$.

                    \blt[grid-aware metrics] Regarding grid-aware metrics, note from Fig.~\ref{fig:usrp_aware16_10_to_100} that RMSE$_\gridnobs$ is generally higher than RMSE$_\gridall$. This is expected, as  the grid points where RMSE$_\gridall$ is evaluated includes the grid points where measurements have been observed. In fact,  as $\obsnum$ increases,  the RMSE$_\gridall$  of any reasonable estimator will continue decreasing
                    until $\obsnum$ reaches $|\measindset|$.
                    At this point, even the trivial estimator that returns
                    $\powest(\loc_{i,j}^\grid)=\powmeas{i,j}$ for all $i,j$ will
                    attain RMSE$_\gridall=0$.

                    \blt[algorithms]
                    Remarkably, the differences across algorithms are not very
                    large. For grid-agnostic metrics, the difference between the
                    best and worst algorithm is around 0.2 dB, whereas for
                    grid-aware metrics, this difference is around 1 dB. As anticipated, \emph{traditional
                        algorithms such as $\neighbornum$-NN and Kriging offer
                        highly competitive performance relative to pure CNN
                        estimators.}

                    %     Thus, depending on the application, using CNN
                    % estimators may not be worth given their need for training on
                    % a large data set.                     

                \end{bullets}

                \begin{figure}[t]
                    \centering
                    \includegraphics[width=\linewidth]{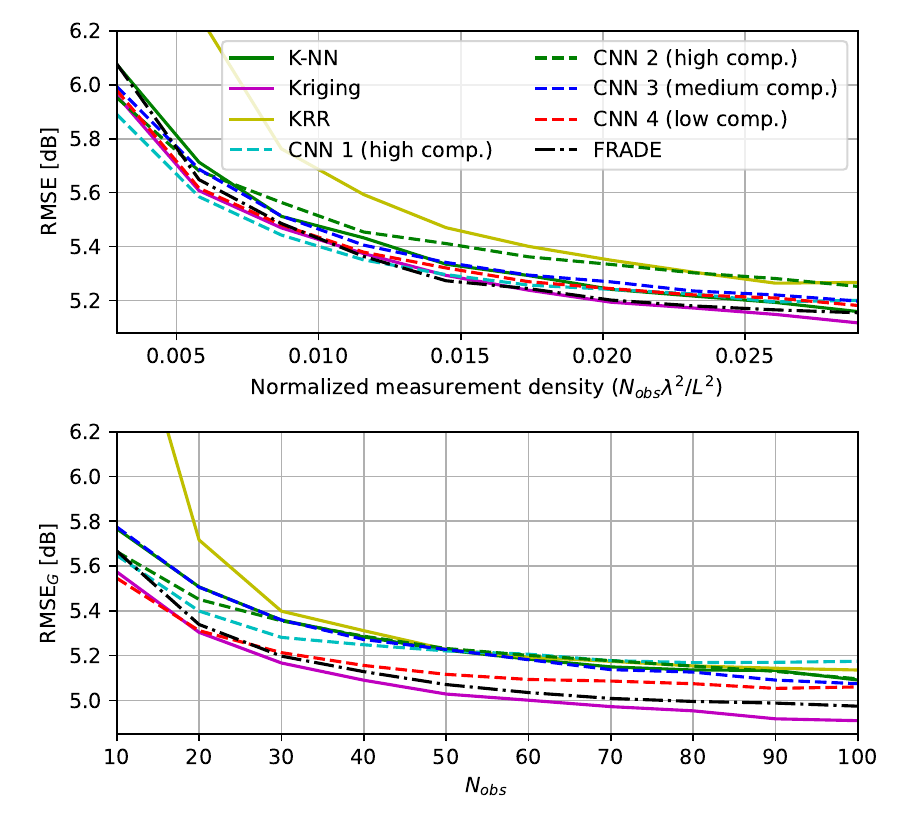}
                    \caption{Grid-agnostic performance metrics for USRP data vs. the number of observations when
                        $\areaside=19.2$~m, $\gridspacing = 1.2$, and the estimators are trained with $\obsnum \in [10, 100]$.}
                    \label{fig:usrp_agnostic16_10_to_100}
                \end{figure}

                \begin{figure}[t]
                    \centering
                    \includegraphics[width=\linewidth]{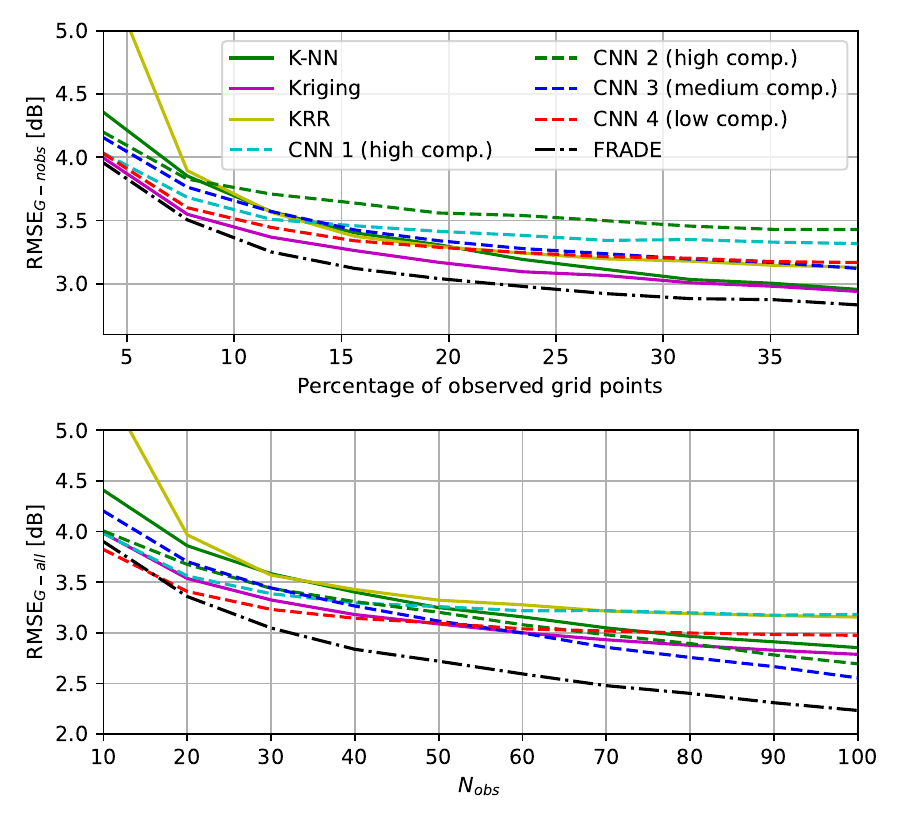}
                    \caption{Grid-aware performance metrics for USRP data vs. the number of observations when
                        $\areaside=19.2$~m, $\gridspacing = 1.2$, and the estimators are trained with $\obsnum \in [10, 100]$.}
                    \label{fig:usrp_aware16_10_to_100}

                \end{figure}
            \end{bullets}

        \end{bullets}

        \blt[vs. num of datasets]
        \begin{bullets}
            \begin{figure}[t]
                \centering
                \includegraphics[width=\linewidth]{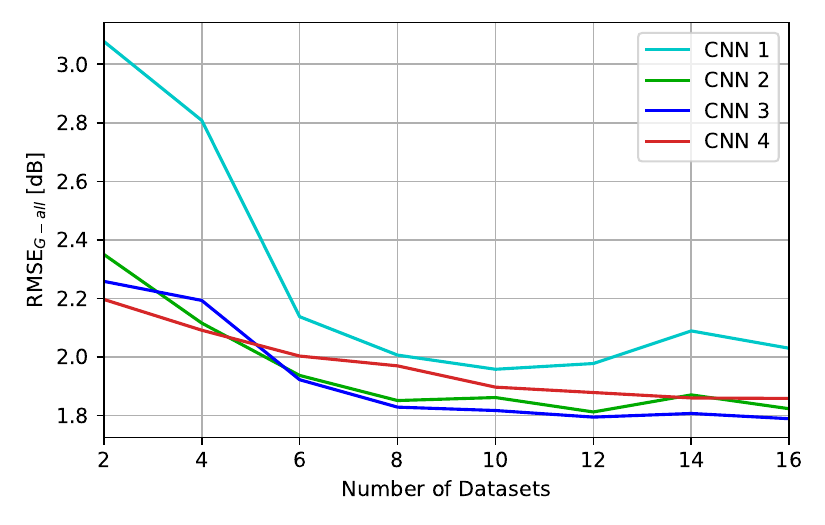}
                \caption{RMSE  for USRP data vs. the number of datasets used during training for the compared CNNs when  $\areaside=19.2$~m, $\gridspacing = 1.2$, and  $\obsnum = 30$. %The test datasets are 10 and 12
                }
                \label{fig:usrp_vs_dataset}
            \end{figure}

            \blt[motivation]Given this unexpected observation, one may suspect that the amount of training data is not sufficient for the CNN estimators.
            \blt[descriptions]To see that this is not the case,
            Fig.~\ref{fig:usrp_vs_dataset} depicts the RMSE$_\gridall$ vs. the number of measurement sets used for training each CNN estimator when $\obsnum = 30$ and $\areaside = 19.2$~m.
            \blt[training and test datasets]Specifically, for each point in the x-axis, the CNN estimators were trained with the first (intense shadowing) measurement sets and tested on the last two measurement sets.
            \blt[training CNNs]At each point, the CNN weights are initialized using the weights obtained using the preceding number of measurement sets.

            \blt[observations]Fig.~\ref{fig:usrp_vs_dataset} demonstrates that  RMSE$_\gridall$ reaches a plateau when  the number of measurement sets used for training is around 10. Thus, the considered training data is sufficient for the previous experiment because additional training data is not expected to significantly alter the performance metrics. The reason why CNN estimators do not offer significantly better performance than traditional estimators is, therefore, not due to a lack of training data, but rather to the fact that CNN estimators regard the RME problem as an image processing problem instead of  exploiting the spatial structure of radio maps; see Sec.~\ref{sec:hybrid}.

        \end{bullets}

        \blt[32 x 32] The previous two experiments adopted a patch size of
        $\areaside=19.2$~m, which resulted in a $16\times 16$ grid. The reason was to ensure that the set of training patches  was sufficiently large and rich to train CNN estimators. In practice, however, one may be interested in constructing radio maps over larger areas. This  is considered in the next experiment, where the patch size is increased to $\areaside=38.4$~m, thereby resulting in an area of 1,475 m$^2$ and a $32\times 32$ grid. The same number of training patches are drawn, but now they have a greater size and, hence, greater overlap. Each patch is thus less informative given the rest.
        % The number of significantly distinct training and testing patches is now smaller, so the statistical significance is reduced.
        \begin{bullets}

            \begin{figure}[t]
                \centering
                \includegraphics[width=\linewidth]{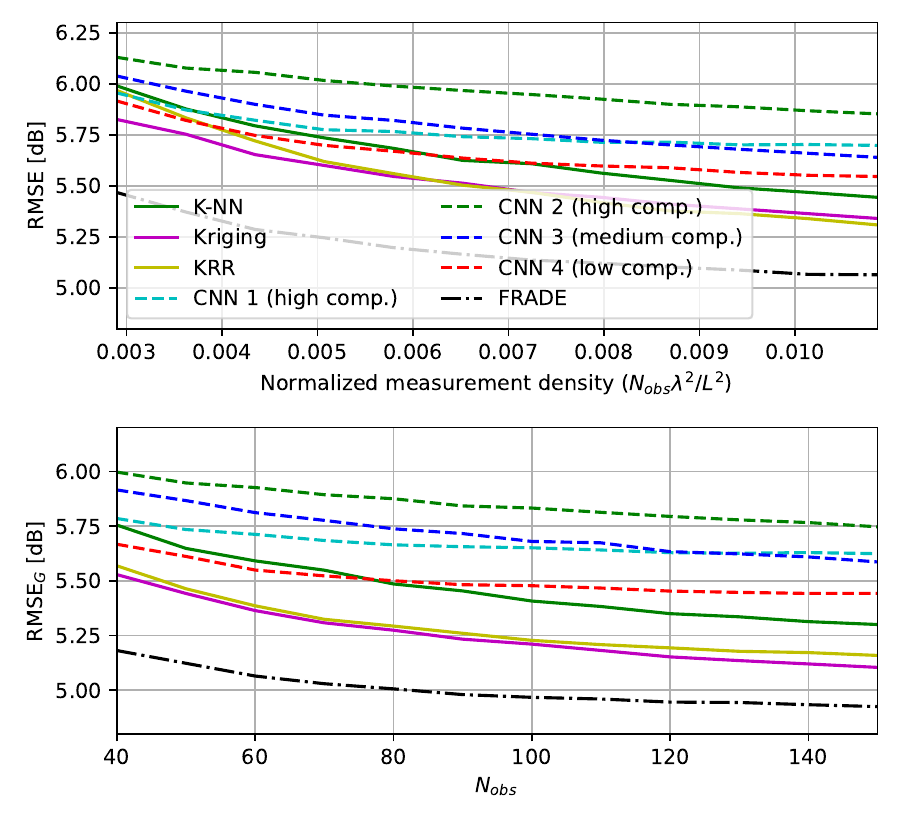}
                \caption{Grid-agnostic
                    performance metrics for USRP data vs. the number of observations when
                    $\areaside=38.4$~m, $\gridspacing = 1.2$, and the estimators are trained with $\obsnum \in [40, 150]$.}
                \label{fig:usrp_agnostic32_40_to_150}
            \end{figure}

            \begin{figure}[t]
                \centering
                \includegraphics[width=\linewidth]{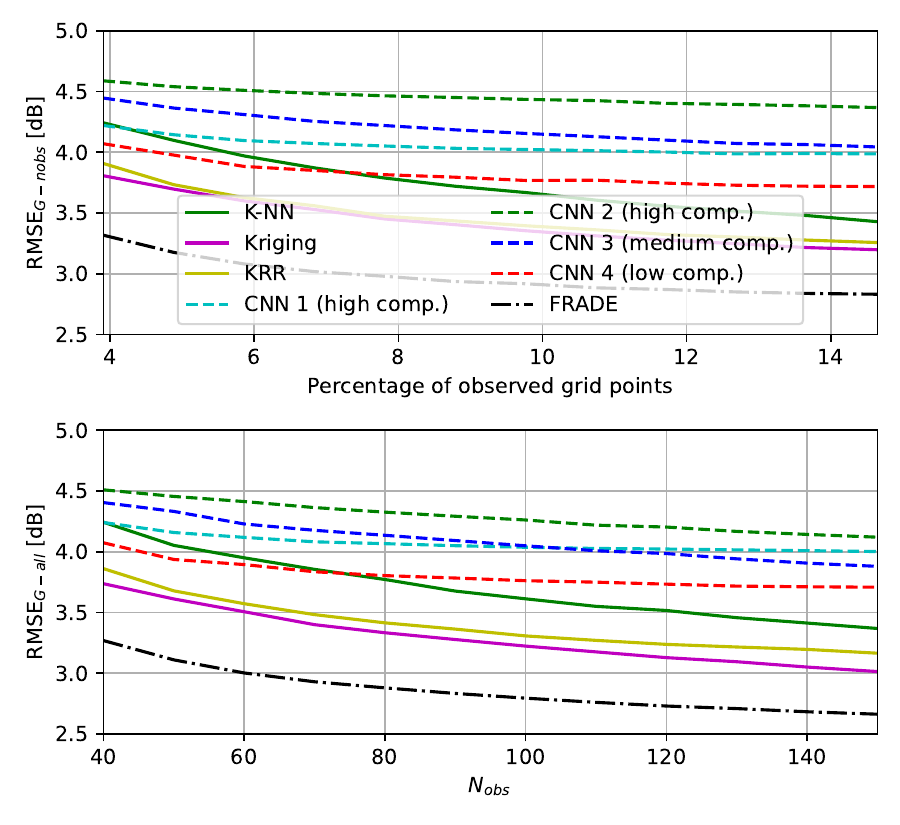}
                \caption{Grid-aware
                    performance metrics for USRP data vs. the number of observations when
                    $\areaside=38.4$~m, $\gridspacing = 1.2$, and the estimators are trained with $\obsnum \in [40, 150]$.}
                \label{fig:usrp_aware32_40_to_150}
            \end{figure}
            \blt[40 to 150 meas]
            \begin{bullets}

                \blt[description]Figs.~\ref{fig:usrp_agnostic32_40_to_150}
                and~\ref{fig:usrp_aware32_40_to_150} show the four  metrics when the estimators were trained by selecting
                $\obsnum$ uniformly at random in $[40, 150]$.
                \blt[observations]
                \begin{bullets}
                    \blt[overall error]For a given $\obsnum$, the error metrics are significantly larger than in Figs.~\ref{fig:usrp_agnostic16_10_to_100} and \ref{fig:usrp_aware16_10_to_100}. This is mainly because the spatial measurement density is now 4 times smaller. For this reason, it makes more sense to compare the metrics at the same normalized measurement density. One can observe, e.g., that all metrics for a density of 0.01 with Kriging are similar to those in Figs.~\ref{fig:usrp_agnostic16_10_to_100} and \ref{fig:usrp_aware16_10_to_100} (check also e.g. 0.005). In contrast, the performance of CNN estimators is significantly degraded since the training dataset is less informative. This is despite the large number of measurements in the dataset; see Sec.~\ref{sec:hybrid}.

                    % Remarkably, the CNNs do not exhibit this behavior: the larger patch size results in greater error due to the reduction in the 
                    % %  This affects both non-CNN and CNN estimators. It also implies that each additional measurement entails a smaller reduction in the estimation error, as can be seen by the smaller slope of the curves in Figs.~\ref{fig:usrp_agnostic32_40_to_150} and \ref{fig:usrp_aware32_40_to_150} compared to Figs.~\ref{fig:usrp_agnostic16_10_to_100} and \ref{fig:usrp_aware16_10_to_100}.
                    % % Second, the 
                    % number of significantly distinct training patches, which reduces the amount of information in the training dataset. 

                    % This affects mainly CNN estimators, which explains their worse performance in Figs.~\ref{fig:usrp_agnostic32_40_to_150} and \ref{fig:usrp_aware32_40_to_150} compared to Figs.~\ref{fig:usrp_agnostic16_10_to_100} and \ref{fig:usrp_aware16_10_to_100}.

                    % The difference in the values of the performance metrics is more pronounced in  Figs.~\ref{fig:usrp_agnostic32_40_to_150} and \ref{fig:usrp_aware32_40_to_150} compared to  Figs.~\ref{fig:usrp_agnostic16_10_to_100} and \ref{fig:usrp_aware16_10_to_100}.
                    %

                    %                    
                    \blt[algorithms]
                    %The performance of the CNN estimators is slightly
                    % degraded with respect to Figs.~\ref{fig:usrp_agnostic16_10_to_100} and
                    % \ref{fig:usrp_aware16_10_to_100} because the larger $\areaside$ dramatically reduces
                    % the number of training patches. 
                    \begin{bullets}%
                        \blt[hybrid]
                        % In contrast, FRADE outperforms the other estimators by a  much wider margin than in Figs.~\ref{fig:usrp_agnostic16_10_to_100} and \ref{fig:usrp_aware16_10_to_100} because its reliance on traditional estimators renders it less sensitive to the amount of information in the training data. Observe, e.g., that 

                        %
                        \blt[autoencoders]Interestingly, the estimators based on autoencoders (CNN~1 and CNN 4) outperform the rest of CNNs,  suggesting that this  architecture is preferable in presence of limited  training data.
                    \end{bullets}%
                \end{bullets}
            \end{bullets}
        \end{bullets}

    \end{bullets}
\end{bullets}
%
%% Figures for USRP
%
%
%
% 32x32 figure USRP
%
%
% rmse vs num dataset usrp
%% Second table
% \usepackage{multirow}

\subsubsection{Experiments with 4G Data}
\label{sec:4gexperiments}
\begin{bullets}
    \blt[Overview]The dataset
    used in the previous experiments was collected  by a custom system that allows the acquisition of a very large number of measurements under highly controlled conditions. This was necessary to investigate the dependence of performance on  the amount of training data.
    In contrast, this section  relies  on measurements of a real-world cellular network to further study  practical aspects of RME.

    % \blt[Data]In the case of the 4G data set, 
    \begin{bullets}

        % \blt[delta] Whenever a grid is constructed, $\gridspacing=4$~m.

        \blt[training/test]
        \begin{bullets}%
            \blt[selected cells]Recall that the 4G dataset comprises  a measurement set for every cell  in two different geographic areas. Since most of the cells are measured only in a small fraction of the $\lggridx\lggridy$ locations, the cell with most measurements is selected in the first area and the two cells with most measurements are selected in the second one. This results in three measurement sets,
            \blt[dataset split]two of them used for training (carrier frequencies of 796 MHz and 952.40 MHz) and the third used for testing (carrier frequency 796 MHz).
            \blt[training]
            \begin{bullets}
                \blt[Number of unique maps]A total of 30,000 estimation instances are formed by first choosing a training measurement set uniformly at random and then randomly choosing a patch; cf. Sec.~\ref{sec:collection}.
                \blt[number of blocks/per patch]Each of these estimation instances is partitioned into an observed and an unobserved set  5 times as described in Sec.~\ref{sec:gridawaremetrics},
                \blt[Total training Examples]resulting in a total of 150,000 training examples.
            \end{bullets}
            \blt[RSRP metric]Due to space limitations, only RSRP is considered. \ifroom{The reason  for selecting  this metric is its greater repeatability, which is a consequence of its lower sensitivity to interference. To corroborate this fact,  Fig.~\ref{fig:histograms_metrics} shows the  histograms of the difference between each measurement and the sample mean of the measurements associated with the same grid point. It is indeed observed that the variance of RSRP is the smallest among the three metrics, which also suggests that RSRP is the most suitable for RME.

                \begin{figure}[t]
                    \centering
                    \includegraphics[width=\linewidth]{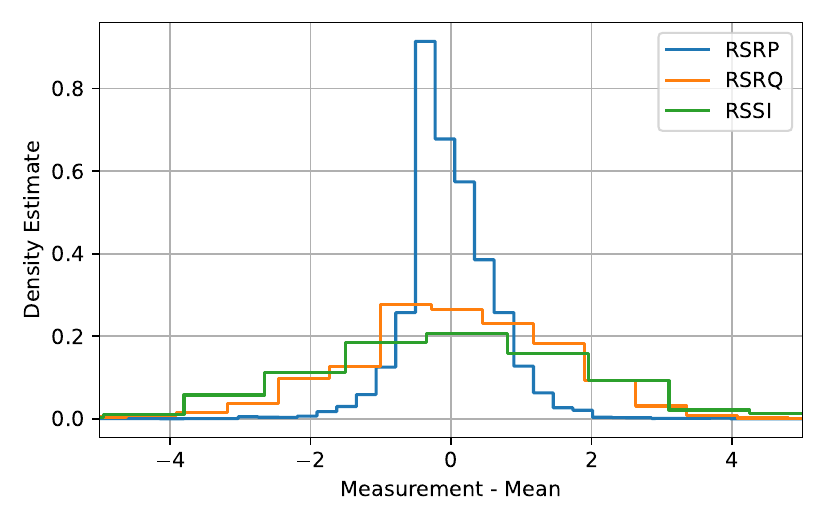}
                    \caption{Histogram plots of the difference between measurements associated with the $(i,j)$-th grid point and their mean for different metrics of one of the  4G measurement sets per metric.}
                    \label{fig:histograms_metrics}
                \end{figure}
            }

        \end{bullets}%

    \end{bullets}

    \blt[figures-Gradiant Combined]

    \begin{bullets}
        %\blt[Histogram of metrics]
        %\begin{bullets}
        %    \blt[reason for choosing rsrp]
        %\end{bullets}
        \blt[16x16]
        \begin{bullets}
            \blt[RSRP 10 to 100 meas]
            \begin{bullets}
                \begin{figure}[t]
                    \centering
                    \includegraphics[width=\linewidth]{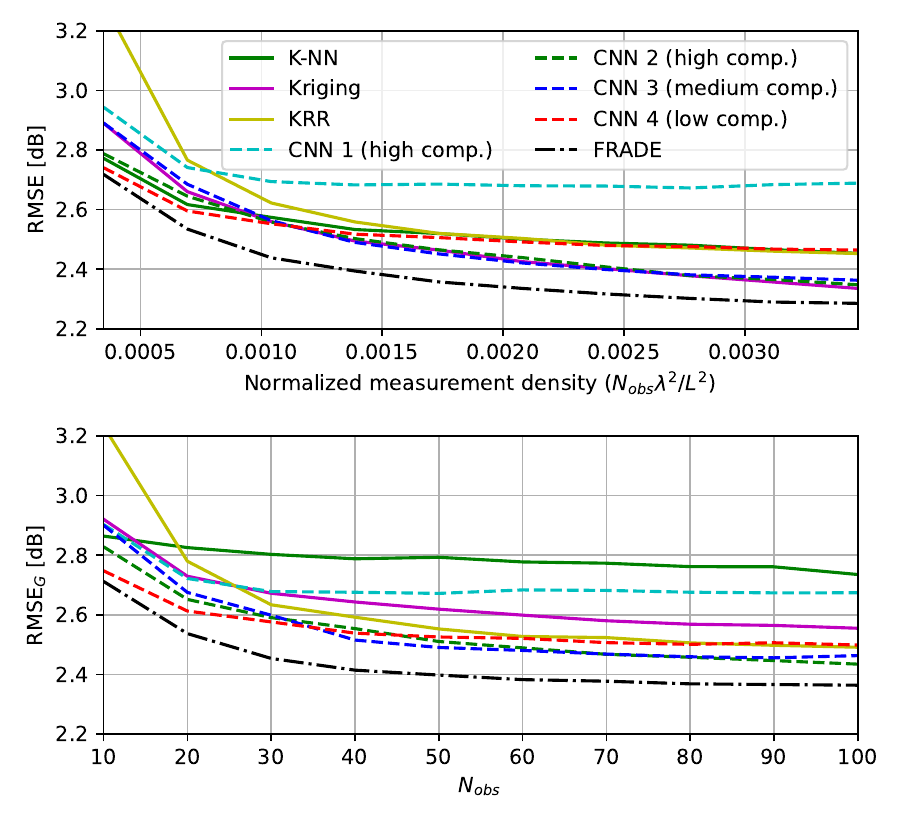}
                    \caption{Grid-agnostic performance metrics for 4G data  vs. the number of observations when
                        $\areaside=64$~m, $\gridspacing = 4$, and the estimators are trained with $\obsnum \in [10, 100]$. }
                    \label{fig:gradiant_agnostic16_10_to_100}
                \end{figure}

                \begin{figure}[t]
                    \centering
                    \includegraphics[width=\linewidth]{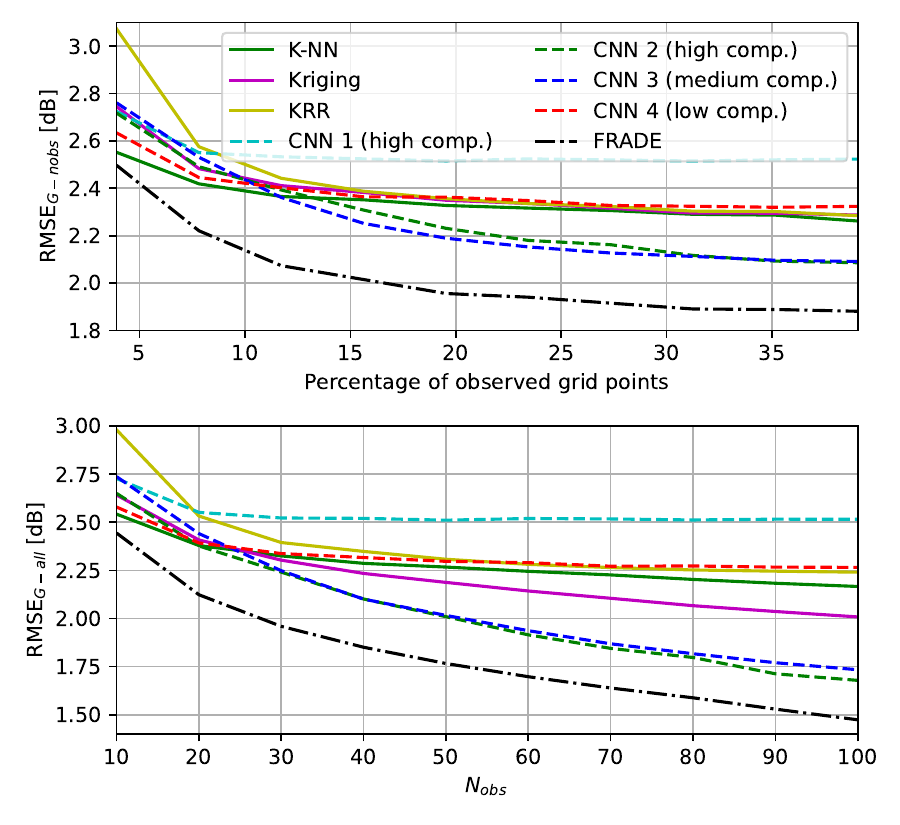}
                    \caption{Grid-aware performance metrics for 4G data  vs. the number of observations when
                        $\areaside=64$~m, $\gridspacing = 4$, and the estimators are trained with $\obsnum \in [10, 100]$.}
                    \label{fig:gradiant_aware16_10_to_100}
                \end{figure}

                \blt[description]Figs.~\ref{fig:gradiant_agnostic16_10_to_100} and \ref{fig:gradiant_aware16_10_to_100} depict the performance metrics
                when the patch side is $\areaside = 64$~m, which results in an area of 4096 m$^2$ and $\gridx=\gridy=\areaside/\gridspacing=16$.
                \blt[observations]%
                \begin{bullets}%
                    \blt[rel. to USRP dataset]First, observe that  the performance metrics are significantly lower than in the USRP dataset. This phenomenon is studied in detail in Sec.~\ref{sec:requiredobservations}.
                    % \blt[grid agnostic metrics]Second, note that RMSE$_\text{G}$ is higher than RMSE. To understand this effect, recall that  5 measurements were acquired  at each grid point. Thus, the selection of the observed measurements used for RMSE$_\text{G}$ (cf. Sec.~\ref{sec:metrics}) yields very accurate information of the true map at a small number of grid points, whereas in the case of RMSE, the information is less accurate but comprises a larger number of grid points. In view of Fig.~\ref{fig:gradiant_agnostic16_10_to_100}, the benefit of  a larger number of grid points dominates. This could not be the case if the measurement error were higher.
                    %
                    \blt[grid aware metrics]Second, the grid-aware metrics take lower values than  RMSE. This is mainly because the measurements were acquired at the grid points. This means that, for a given $\obsnum$, the number of actual observations for grid-aware metrics is 5 times greater than in the case of RMSE.

                    \blt[performance analysis of algorithms]Although  differences are small, Kriging is now outperformed by some CNN estimators in 3 out of the 4 metrics. In the case of K-NN, RMSE$_\text{G}$ is poor because many neighbors are selected from the same grid point, which reduces the spatial resolution of its estimates. Averaging nearby measurements may be a desirable pre-processing step when applying this algorithm. This is further corroborated by the remarkably good values that this estimator achieves for the grid-aware metrics when $\obsnum$ is low.

                \end{bullets}
            \end{bullets}
        \end{bullets}

    \end{bullets}
\end{bullets}

\subsection{The Complexity of the RME Problem}
\label{sec:requiredobservations}
\begin{bullets}%
    \blt[overview]Observe that the considered estimators generally offer a better performance on the 4G dataset than on the USRP dataset, despite the fact that the amount of training data in the latter is significantly greater than in the former.
    \blt[complexity]The reason is that estimating all radio maps is not equally difficult. As intuition predicts, for a given number of observations, the estimation error is generally smaller for families of radio maps with smaller variability.

    \blt[free space]
    \begin{bullets}
        \blt[theoretical]The influence of this variability on estimation performance was analyzed mathematically for free-space propagation in~\cite{romero2024theoretical}. In short, the estimation error was shown to decrease under general conditions with the distance to the transmitters.
        \blt[empirical]To empirically verify this  phenomenon, the   LOS
        measurement sets of the USRP dataset were used to obtain 
        Fig.~\ref{fig:rmse_vs_dist}, which depicts the RMSE of Kriging vs.  the distance between 
        the transmitter and the center of the mapped region.
        As expected,  the error
        exhibits a pronounced decreasing trend with respect to the  distance to the transmitter, which
        corroborates the analysis in~\cite{romero2024theoretical}. The variability around the mean is due to the reflections on the ground, trees, and mountains near the measurement region.

        \begin{figure}
            \centering
            \includegraphics[width=\linewidth]{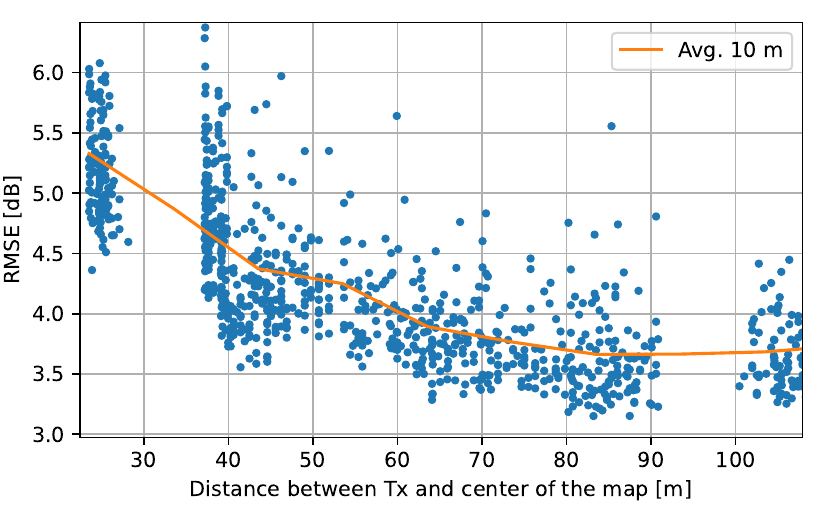}
            \caption{
                    Estimation error as a function of the distance to the transmitter for the USRP data. The Kriging estimator is used to obtain power estimates. $\areaside=43.2$~m, $\gridspacing = 1.2$, $\obsnum\lambda^2/\areaside^2=0.001$.                                    }
            \label{fig:rmse_vs_dist}
        \end{figure}

    \end{bullets}

    \blt[non free space]
    \begin{bullets}
        \blt[factors]Even in scenarios without  free-space propagation or LOS, this phenomenon still holds in many cases~\cite{romero2024theoretical}. But in such scenarios, another critical factor is shadowing.
        \blt[4G dataset]Both distance and shadowing  explain  why the estimation performance was seen to be better in the 4G dataset than in the USRP dataset:  the transmitters in the 4G dataset are farther away from the mapped region and, besides, the shadowing patters induced by the metallic reflectors of the USRP dataset are considerably more abrupt than the ones produced by the mountains and other terrain features affecting propagation in the 4G dataset. This smaller variability in the 4G dataset is also consistent  with the fact that the performance of K-NN is less sensitive to $\obsnum$ in this dataset than in the USRP dataset.

        \blt[experiment]
        \begin{bullets}
            \blt[data]To study the impact of propagation effects such as shadowing on estimation performance, the data is partitioned into scenarios based on the nature of the dominating propagation phenomena. Recall that the USRP dataset is already divided into an \emph{ intense shadowing scenario}, which contains the measurement sets with reflectors, and  a \emph{LOS scenario}, which contains the rest. The latter is further subdivided into two categories based on the distance to the transmitter. On the other hand, the 4G dataset makes up the \emph{smooth shadowing  scenario}, where shadowing evolves slowly across space since it is caused by  distant mountains and other terrain features.

            \blt[figure]Fig.~\ref{fig:rmse_vs_samp_density-kriging} plots the estimation  error vs. the measurement density for each scenario. The effects of distance and shadowing are here manifest: the error is certainly larger for smaller distances and more abrupt shadowing. In addition, as intuition predicts, the error is seen to be less sensitive to the measurement density when spatial variability is smaller. Finally, note that this figure is also useful to draw quantitative (but approximate) conclusions such as ``with 1 observation per area of 1,000 square wavelengths, the error is around 5.2 dB when the transmitter is a few tens of wavelengths away and around 3 dB when the transmitter is at least a few hundreds of wavelengths away." Thus, this figure may guide the practitioner to determine whether RME yields a sufficiently low error for a given application.

            \begin{figure}
                \centering
                \includegraphics[width=\linewidth]{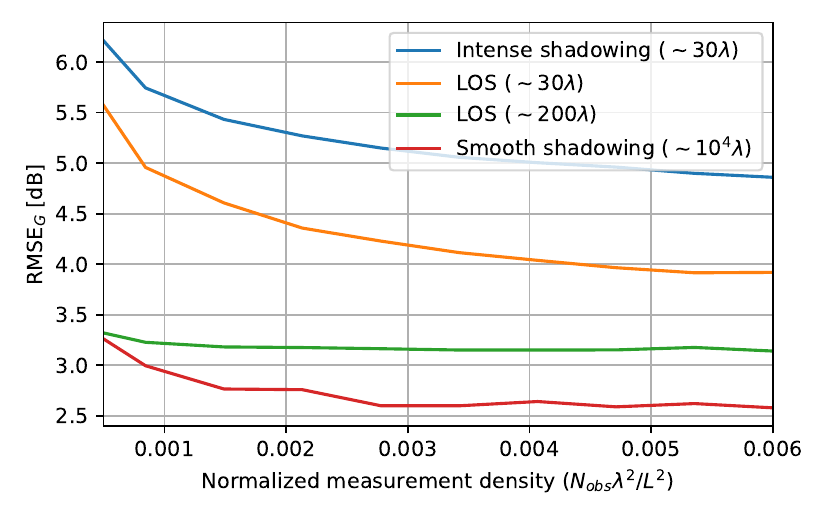}
                \caption{                    
                        Estimation error in each propagation scenario. The distance to the transmitter is in the legend. The Kriging estimator is used, where the values of $\ushadvar$ and $\dist$ are chosen to approximately minimize RMSE on a separate training set for each value of  $\obsnum$ and scenario. $\areaside=40$~m, and $\gridspacing = 4$.                    
                }
                \label{fig:rmse_vs_samp_density-kriging}
            \end{figure}

        \end{bullets}

    \end{bullets}

    % %
    % \blt[factors that affect error]To this end, note that the factors determining the estimation error are:
    % \begin{bullets}%
    %     \blt (F1) the spatial structure of the true map, dictated by  Maxwell's equations, the  environment, and the transmitters;
    %     %
    %     \blt (F2) the number of measurements and their locations;
    %     %
    %     \blt (F3) the measurement error, caused by noise,
    %     interference, and the limited number of time samples used to measure power
    %     at a given location;
    %     %
    %     \blt (F4) the localization error;
    %     %
    %     \blt (F5) the estimator and its parameters; and
    %     %
    %     \blt (F6) the considered performance metric.
    % \end{bullets}%
    % Thus, it is not possible to establish precise claims on the number of observations required to achieve a target estimation error in general. However, in the controlled conditions of the experiments here, factors (F3) to (F6) are fixed and one can explore the influence of (F1) and (F2).

    % \blt[outline]
    % \begin{bullets}%
    %     \blt[F2]For (F2), the estimation error can be computed for different number of observations and sampling modes.
    %     %
    %     \blt[F1]For (F1), 
    % \end{bullets}%

\end{bullets}%    

\subsection{Hybrid Estimators}%
\label{sec:hybrid}

\begin{bullets}
    \blt[Failure of CNNs]The experiments from Sec.~\ref{sec:performance} revealed that existing CNN estimators did not significantly outperform  traditional approaches despite being trained with measurements at hundreds of thousands of locations.
    \blt[decomposition]    
    To investigate the cause of such a learning inefficiency, it is instructive to think of a radio map as a combination of a somehow \emph{spatially smooth (parsimonious) component} (which includes free-space loss and shadowing caused by distant obstacles) and a \emph{deviation} from this component.
    \blt[traditional]By design, traditional (function regression) schemes learn to estimate this \emph{smooth component} from a very small amount of training data since they condense the variability of this component in just  a few parameters (e.g. $\ushadvar$ and $\dist$ in the case of Kriging). In machine learning terminology, traditional estimators have an \emph{inductive bias} that favors spatially smooth estimates. 
    \blt[cnn]This is not the case of CNN estimators, which can learn highly complex patterns at the expense of millions of parameters and a giant amount of data. These estimators are agnostic of the spatial structure of radio maps: they adopt an \emph{image processing approach} by which \emph{they learn both the smooth and deviation components from data} and this is a core reason for the observed learning inefficiency.

    \blt[hypothesis verification \ra combination]To verify this hypothesis, a new family of \emph{hybrid estimators} is considered next. The idea is to \emph{use a traditional estimator   to learn the smooth component and a CNN estimator to learn the deviation component}. In this way, if the amount of training data is small, performance will be reasonable since the smooth component will be learned. If it is large, the deviation components will also be learned, thereby improving the performance relative to traditional estimators.

    \blt[design]There are many possibilities for constructing hybrid estimators. One can, for example, consider ensemble or mixture-of-experts approaches~\cite[Sec. 14.5.3]{bishop2006}, which are popular  in machine learning. Since the purpose of this work is not to develop new estimators but  to study the RME problem and evaluate existing schemes from a \emph{neutral standpoint}, one of the arguably simplest hybrid estimators is considered, leaving more sophisticated alternatives for future work.

    \blt[frade]This estimator, referred to as \emph{function regressor-assisted deep estimator (FRADE)}, is a modification of CNN 3 that obtains a map estimate as follows.
    First, three estimates of the map on the grid are obtained using the three estimators in Sec.~\ref{sec:sim:nonCNNestimators} and
    concatenated to $\powmeasmat$ and $\samplingmask$ to form a $5 \times
        N_y \times N_x$ tensor. This is passed to a network with the same architecture as CNN 3, except for the input layer, where the kernels need a greater size to accommodate the larger dimensionality of the input tensor. This network
    produces the $N_y \times N_x$ estimate $\powestmat$.
    The values of the parameters of the traditional estimators are the ones obtained by training each of them separately; see Sec.~\ref{sec:sim:nonCNNestimators}.
    To improve performance even further, these estimators take  the plain measurements as input rather than their grid-quantized version.

    \blt[performance]A curve for FRADE was included in the figures of Sec.~\ref{sec:performance}. This simple improvement over CNN 3 is seen to outperform both traditional and CNN estimators in most cases. For example, in Fig.~\ref{fig:usrp_aware32_40_to_150},
    FRADE requires 90 measurements less  than the best competitor to attain
    RMSE$_\gridall$
    $\approx$ 3 dB. Note, however, that FRADE need not result in the best values for absolutely all metrics since it is trained specifically for RMSE$_\gridall$.     
    \blt[conclusion]In any case, the results in these figures  verify the aforementioned hypothesis relative to the learning inefficiency of plain CNN estimators and opens the door for the development of more sophisticated hybrid schemes. 

\end{bullets}

\section{Closing Remarks}
\label{sec:closingremarks}

\subsection{Summary of the Key Observations}
\label{sec:keyobservations}
\begin{bullets}
    \blt[overview]The main findings  from the experiments in Sec.~\ref{sec:experiments} are summarized next:

    \blt[Key points]
    \begin{enumerate}
        \item                  A reasonably small  estimation error is observed in a wide variety of setups, which suggests that the estimated maps may be sufficiently accurate for real-world applications and, therefore, constitutes empirical evidence for the practical viability of RME.

        \item Grid-aware metrics are typically lower than grid-agnostic metrics, which suggests that averaging out small-scale fading facilitates RME.

        \item Among the considered alternatives, obtaining the grid-quantized
              measurement $\powmeas{i,j}$ by averaging the measurements assigned to
              $\loc^{\grid}_{i, j}$ in dB units   seems to be the most effective approach to average out small-scale fading; cf. Sec.~\ref{sec:usrpdsdc}.

              \item The difficulty of the RME problem is  mainly determined by (i) the distance to the transmitter, as predicted theoretically in~\cite{romero2024theoretical}, and (ii) the shadowing profile. 

              \newlength{\mycolumnwidth}
              \setlength{\mycolumnwidth}{7.7cm}
              \begin{table*}[]
                  \centering
                  \caption{Table  summarizing the relative performance of the estimators in the literature.                      
                          When the number of measurements is said to be \emph{high}, it means
                          that there exists an $\obsnum'$ such that the corresponding estimator outperforms the others whenever $\obsnum\geq \obsnum'$. Similarly for \emph{low}.                      
                      FRADE is excluded from the comparison since it  outperforms existing estimators in most cases.}
                  \label{tab:summary}
                  \begin{tabular}{|c|p{\mycolumnwidth}|cccc|}
                    \hline
                    \multirow{2}{*}{\textbf{Estimators}} & \multicolumn{1}{c|}{\multirow{2}{*}{\textbf{\begin{tabular}[c]{@{}c@{}}Strengths/Limitations\end{tabular}}}}                                                                                                                                                                                                                & \multicolumn{4}{c|}{\textbf{Best Performance Cases}}                                                                                                                                            \\ \cline{3-6}
                                                        & \multicolumn{1}{c|}{}                                                                                                                                                                                                                                                                                                       & \multicolumn{1}{c|}{\textbf{Metric}}                 & \multicolumn{1}{c|}{\textbf{\#Meas.}}            & \multicolumn{1}{c|}{\textbf{Dataset}}               & \textbf{$\gridy \times \gridx$} \\ 
                    \hline
                    \multirow{3}{*}{\textbf{K-NN}}       & \multirow{3}{*}{\begin{tabular}[c]{@{}p{\mycolumnwidth}@{}}Strengths: 1) Simple and easy to implement. 2) Requires a small amount of training data.\\ Limitations:                                                            1) Limited ability to  capture complex spatial patterns.\end{tabular}}                                       & \multicolumn{1}{c|}{RMSE$_\gridnobs$}                & \multicolumn{1}{c|}{High}                        & \multicolumn{1}{c|}{USRP, $\obsnum\in[10, 100]$}    & $16\times16$                    \\ \cline{3-6}
                                                        &                                                                                                                                                                                                                                                                                                                             & \multicolumn{1}{c|}{RMSE$_\gridnobs$}                & \multicolumn{1}{c|}{Low}                         & \multicolumn{1}{c|}{4G Data, $\obsnum\in[10, 100]$} & $16\times16$                    \\ \cline{3-6}
                                                        &                                                                                                                                                                                                                                                                                                                             & \multicolumn{1}{c|}{RMSE$_\gridall$}                 & \multicolumn{1}{c|}{Low}                         & \multicolumn{1}{c|}{4G Data, $\obsnum\in[10, 100]$} & $16\times16$                    \\ 
                    \hline
                    \multirow{12}{*}{\textbf{Kriging}}   & \multirow{12}{*}{\begin{tabular}[c]{@{}p{\mycolumnwidth}@{}}Strengths: 1)  Requires a small amount of training data. 2) Offers uncertainty estimates for predictions.\\ Limitations:1) Computationally intensive for large $\obsnum$ due to matrix inversion.\end{tabular}}                                                 & \multicolumn{1}{c|}{RMSE$_\gridnobs$}                & \multicolumn{1}{c|}{Low}                         & \multicolumn{1}{c|}{USRP, $\obsnum\in[10, 100]$}    & $16\times16$                    \\ \cline{3-6}
                                                        &                                                                                                                                                                                                                                                                                                                             & \multicolumn{1}{c|}{RMSE}                            & \multicolumn{1}{c|}{High}                        & \multicolumn{1}{c|}{USRP, $\obsnum\in[10, 100]$}    & $16\times16$                    \\ \cline{3-6}
                                                        &                                                                                                                                                                                                                                                                                                                             & \multicolumn{1}{c|}{RMSE$_\text{G}$}                 & \multicolumn{1}{c|}{High}                        & \multicolumn{1}{c|}{USRP, $\obsnum\in[10, 100]$}    & $16\times16$                    \\ \cline{3-6}
                                                        &                                                                                                                                                                                                                                                                                                                             & \multicolumn{1}{c|}{RMSE$_\gridnobs$}                & \multicolumn{1}{c|}{High}                        & \multicolumn{1}{c|}{USRP, $\obsnum=50$}             & $16\times16$                    \\ \cline{3-6}
                                                        &                                                                                                                                                                                                                                                                                                                             & \multicolumn{1}{c|}{RMSE}                            & \multicolumn{1}{c|}{High}                        & \multicolumn{1}{c|}{USRP, $\obsnum=50$}             & $16\times16$                    \\ \cline{3-6}
                                                        &                                                                                                                                                                                                                                                                                                                             & \multicolumn{1}{c|}{RMSE$_\text{G}$}                 & \multicolumn{1}{c|}{High}                        & \multicolumn{1}{c|}{USRP, $\obsnum=50$}             & $16\times16$                    \\ \cline{3-6}
                                                        &                                                                                                                                                                                                                                                                                                                             & \multicolumn{1}{c|}{RMSE$_\gridall$}                 & \multicolumn{1}{c|}{Low}                         & \multicolumn{1}{c|}{USRP, $\obsnum=50$}             & $16\times16$                    \\ \cline{3-6}
                                                        &                                                                                                                                                                                                                                                                                                                             & \multicolumn{1}{c|}{RMSE$_\gridnobs$}                & \multicolumn{1}{c|}{All}                         & \multicolumn{1}{c|}{USRP, $\obsnum\in[40, 150]$}    & $32\times32$                    \\ \cline{3-6}
                                                        &                                                                                                                                                                                                                                                                                                                             & \multicolumn{1}{c|}{RMSE$_\gridall$}                 & \multicolumn{1}{c|}{All}                         & \multicolumn{1}{c|}{USRP, $\obsnum\in[40, 150]$}    & $32\times32$                    \\ \cline{3-6}
                                                        &                                                                                                                                                                                                                                                                                                                             & \multicolumn{1}{c|}{RMSE}                            & \multicolumn{1}{c|}{Low}                         & \multicolumn{1}{c|}{USRP, $\obsnum\in[40, 150]$}    & $32\times32$                    \\ \cline{3-6}
                                                        &                                                                                                                                                                                                                                                                                                                             & \multicolumn{1}{c|}{RMSE$_\text{G}$}                 & \multicolumn{1}{c|}{All}                         & \multicolumn{1}{c|}{USRP, $\obsnum\in[40, 150]$}    & $32\times32$                    \\ \cline{3-6}
                                                        &                                                                                                                                                                                                                                                                                                                             & \multicolumn{1}{c|}{RMSE}                            & \multicolumn{1}{c|}{High}                        & \multicolumn{1}{c|}{4G Data, $\obsnum\in[10, 100]$} & $16\times16$                    \\ \hline
                    \textbf{KRR}                         & \begin{tabular}[c]{@{}p{\mycolumnwidth}@{}}Strengths:                             1) Requires a small amount of training data.                         2) Theoretical performance guarantees~\cite{romero2017spectrummaps}. \\ Limitations: 1) Computationally intensive for large $\obsnum$ due to matrix inversion.
                                                            \end{tabular}       & \multicolumn{1}{c|}{RMSE}                            & \multicolumn{1}{c|}{High}                        & \multicolumn{1}{c|}{USRP, $\obsnum\in[40, 150]$}    & $32\times32$                                                                                                                                                                                                          \\ 
                    \hline
                    \textbf{DNN 1}                       & \begin{tabular}[c]{@{}p{\mycolumnwidth}@{}}Strengths: 1) Provides uncertainty metrics to enable active sensing~\cite{shrestha2022surveying}.\\ Limitations: 1) Computationally expensive due to the large number of parameters.\end{tabular}                                                                                & \multicolumn{1}{c|}{RMSE}                            & \multicolumn{1}{c|}{Low}                         & \multicolumn{1}{c|}{USRP, $\obsnum\in[10, 100]$}    & $16\times16$                    \\ 
                    \hline
                    \multirow{4}{*}{\textbf{DNN 2}}      & \multirow{4}{*}{\begin{tabular}[c]{@{}p{\mycolumnwidth}@{}}Strengths: 1) Ability to capture complex patterns if sufficient data is given.\\ Limitations: 1) More trainable parameters than other DNN estimators.\end{tabular}}                                                                                                               & \multicolumn{1}{c|}{\multirow{2}{*}{RMSE$_\gridnobs$}}                
                    & \multicolumn{1}{c|}{\multirow{2}{*}{High}}                        & \multicolumn{1}{c|}{\multirow{2}{*}{4G Data, $\obsnum\in[10, 100]$}} & \multicolumn{1}{c|}{\multirow{2}{*}{$16\times16$}}                    
                    \\[1.1em]
                                                        
                    \cline{3-6}
                                                        &                                                                 & \multicolumn{1}{c|}{\multirow{2}{*}{RMSE$_\text{G}$}}                 & \multicolumn{1}{c|}{\multirow{2}{*}{High}}                        & \multicolumn{1}{c|}{\multirow{2}{*}{4G Data, $\obsnum\in[10, 100]$}} & \multirow{2}{*}{$16\times16$}                    \\[1.1em]
                    \hline
                    \multirow{5}{*}{\textbf{DNN 3}}      & \multirow{5}{*}{\begin{tabular}[c]{@{}p{\mycolumnwidth}@{}}Strengths:                                      1) Ability to capture complex patterns if sufficient data is given. 2) Provides uncertainty metrics to enable active sensing.                    3) Similar overall performance to DNN 2 with a smaller number of parameters. \\
                                                                Limitations:               1) More trainable parameters than DNN 4.
                                                            \end{tabular}}
                                                        & \multicolumn{1}{c|}{\multirow{2.5}{*}{RMSE$_\gridall$}}                                                                                                                                                                                                                                                                                        & \multicolumn{1}{c|}{\multirow{2.5}{*}{High}}                            & \multicolumn{1}{c|}{\multirow{2.5}{*}{USRP, $\obsnum\in[10, 100]$}} & \multirow{2.5}{*}{$16\times16$}                                                                          \\[1.6em] \cline{3-6}
                                                        &                                                                                                                                                                                                                                                                                                                             & \multicolumn{1}{c|}{\multirow{2.5}{*}{RMSE$_\gridall$}}                 & \multicolumn{1}{c|}{\multirow{2.5}{*}{High}}                        & \multicolumn{1}{c|}{\multirow{2.5}{*}{USRP, $\obsnum=50$}}             & \multirow{2.5}{*}{$16\times16$}                    \\ [1.6em]\hline
                    \multirow{6}{*}{\textbf{DNN 4}}      & \multirow{6}{*}{\begin{tabular}[c]{@{}p{\mycolumnwidth}@{}}Strengths:                                        1) Lowest complexity among all DNN estimators.                                 2) Strong performance when the amount of training data or observations is low. \\
                                                                                    Limitations:                                                             1) Performance may be sensitive to the value of $\obsnum$ used for training.
                                                                                \end{tabular}}                                               & \multicolumn{1}{c|}{RMSE$_\text{G}$}                 & \multicolumn{1}{c|}{Low}                         & \multicolumn{1}{c|}{USRP, $\obsnum\in[10, 100]$}    & $16\times16$                                                                                                                                                                                                                                                    \\ \cline{3-6}
                                                        &                                                                                                                                                                                                                                                                                                                             & \multicolumn{1}{c|}{RMSE$_\gridall$}                 & \multicolumn{1}{c|}{Low}                         & \multicolumn{1}{c|}{USRP, $\obsnum\in[10, 100]$}    & $16\times16$                    \\ \cline{3-6}
                                                        &                                                                                                                                                                                                                                                                                                                             & \multicolumn{1}{c|}{RMSE}                            & \multicolumn{1}{c|}{Low}                         & \multicolumn{1}{c|}{USRP, $\obsnum=50$}             & $16\times16$                    \\ \cline{3-6}
                                                        &                                                                                                                                                                                                                                                                                                                             & \multicolumn{1}{c|}{RMSE$_\text{G}$}                 & \multicolumn{1}{c|}{Low}                         & \multicolumn{1}{c|}{USRP, $\obsnum=50$}             & $16\times16$                    \\ \cline{3-6}
                                                        &                                                                                                                                                                                                                                                                                                                             & \multicolumn{1}{c|}{RMSE}                            & \multicolumn{1}{c|}{Low}                         & \multicolumn{1}{c|}{4G Data, $\obsnum\in[10, 100]$} & $16\times16$                    \\ \cline{3-6}
                                                        &                                                                                                                                                                                                                                                                                                                             & \multicolumn{1}{c|}{RMSE$_\text{G}$}                 & \multicolumn{1}{c|}{Low}                         & \multicolumn{1}{c|}{4G Data, $\obsnum\in[10, 100]$} & $16\times16$                    \\ \hline
                \end{tabular}
              \end{table*}

        \item Among the estimators in the literature, no one outperforms the others in all scenarios. Each one exhibits the best performance in certain situations; cf. Table~\ref{tab:summary}.

        \item Traditional estimators exhibit noteworthy good performance.  The considered Kriging algorithm, for example, proved highly versatile with top performance in a wide variety of cases, which provides further empirical evidence for the correlated shadowing model in \cite{gudmundson1991correlation}.
             
        \item Even when there is sufficient training data, the performance gains of using CNN estimators does not exceed a fraction of a dB. Besides, as observed in Sec.~\ref{sec:usrpexperiments} (see also Sec.~\ref{sec:collection}), large patch sizes greatly increase the data needs of  CNN estimators, which renders traditional estimators more appealing for  large areas.
        
        \item The learning inefficiency of existing CNN estimators is explained by the absence of a suitable inductive bias. To verify this hypothesis, a new family of \emph{hybrid estimators} was considered. One of the simplest members of this family, named FRADE, was seen to outperform all existing estimators. Thus, hybrid estimators effectively counteract the learning inefficiency of CNN schemes and opens the door to further performance improvements by  developing more sophisticated hybrid estimators. 

        \item U-Net architectures tend to manifest superior performance when sufficient training data and measurements ($\obsnum$) are given. Autoencoder architectures tend to be preferable otherwise. 

        \item Comparing RMSE and RMSE$_\text{G}$ indicates that some estimators are sensitive to how clustered  the measurement  locations are.

              \nextversion{
        \item Depending on the dataset and estimator, it may be worth using different parameter values depending on the number of observations
              ($\obsnum$).
              }

        % \item RSRP is  more suitable than RSRQ and RSSI for RME due to its greater repeatability.

    \end{enumerate}
\end{bullets}

\subsection{Limitations}
\label{sec:discussion}
\begin{bullets}%
    \blt[intro]As every empirical study, the work at hand features  limitations, which include
    %
    % \blt[strengths of the study]Among the strengths, one can highlight the following:
    % \begin{bullets}%
    %     \blt[large amount of data](i) The collected dataset is significantly larger than in all previous works and comprises multiple frequency bands, especially in the context of cellular communications.
    %     %
    %     \blt[large number of estimators](ii) A wide range of existing estimators were compared. Most of them had never been compared with each other.
    %     % Remarkably, several CNN estimators had not been compared with traditional estimators, especially when the parameters of the latter are trained.
    %     %
    %     \blt[several perf. metrics](iii) The four considered performance metrics are useful to capture the needs of a wide variety of applications and illustrate different effects and trade-offs in RME.
    % \end{bullets}%
    %
    \blt[limitations of the study]the following:
    \begin{bullets}%
        \blt[limited to power maps](i) Only the estimation of power-related metrics was considered. Different kinds of radio maps (see e.g.~\cite{romero2022cartography}) will require separate studies.
        \blt[limited setups](ii) Although the collected data spans a significant collection of practically relevant scenarios, new data must be collected for empirically analyzing estimators in other setups, such as those involving higher frequency bands (e.g. mmWave bands), urban scenarios, and user mobility. In particular, CNN estimators can exploit side information on obstacles such as buildings. This has not been considered in the present work. 
        \blt[limited collection of estimators](iii) Some estimators in the literature were  not analyzed here. To allow researchers to assess the performance of those estimators, our datasets and code will be published.
        \blt[kriging](iv) The only version of Kriging  considered here is the model-based MMSE estimator of \cite{shrestha2022surveying}.
            %  Despite it was applied to  scenarios that differ from the one for which this estimator was conceived, its performance was one of the best. However, other (e.g. model-agnostic) versions of Kriging should be considered in future work.        
        %
        \blt[antenna directivity](v) Although the heading of the UAV was kept constant to reduce the impact of antenna directivity, the impact of this factor needs to be further investigated. 

    \end{bullets}%

\end{bullets}%

\subsection{Conclusions}
\label{sec:conclusions}
\begin{bullets}
    \blt[overview] This work presented the first comprehensive, rigorous, and reproducible study of RME with real data. 
    \blt[experiment goal] Real-data experiments involving a wide range of estimators, performance metrics, and estimation scenarios were conducted to study the viability of RME in practice, the complexity and special features of the RME problem, and the performance of existing estimators. 
    \blt[observations]  The main  observations were summarized in Sec.~\ref{sec:keyobservations}.
    \blt[DNN]Remarkably, CNN estimators were seen to suffer from a certain learning inefficiency, which emerges from the fact that these estimators are conceived more for image processing than for estimating spatial fields. This effect was counteracted by hybrid estimators, which combine traditional and CNN schemes.
    
    \blt[choice of estimator]The quantitative results here can assist researchers and practitioners in determining the estimation error that can be expected in a practical scenario and in selecting a suitable estimator.  
    The key factors in this  decision were identified and analyzed. See Sec.~\ref{sec:keyobservations} and 
     Table~\ref{tab:summary}.
    \blt[reproducibility]The vast amount of data collected for this study is also published along with the simulator needed to reproduce the results. 
    %
    % \blt[collected data] Two large datasets were collected with two different measurement systems, one of them specifically developed for this work.
    % %
    % \blt[metrics]Performance was evaluated using four different metrics
    %    

    % \blt[Proposed estimator]\begin{changes2}
    %     Hybrid estimators (like FRADE)  are expected outperform traditional and CNN-based estimators in most cases. This
    %     calls for research on estimators of this kind.
    % \end{changes2}

    \blt[future directions]Future directions include  addressing the limitations listed in Sec.~\ref{sec:discussion}, developing more sophisticated hybrid estimators, and constructing other kinds of maps such as power spectral density maps, outage maps, and channel-gain maps~\cite{romero2022cartography}.

    % \blt[limitations of the study]Among the limitations of the present
    % study, one may highlight the following:
    % \begin{bullets}
    %     \blt[size of the dataset] 1) although the number of measurement
    %     locations in the dataset is by far the largest in the literature, it
    %     is still insufficient to train the ``data-hungry'' CNN estimators. A
    %     larger performance gain relative to traditional estimators must be
    %     expected when CNN estimators are trained on more data.
    %     %
    %     \blt[same parameters] 2) All estimators used the same parameters in
    %     all experiments. Performance improvements may be expected if they are
    %     adjusted depending on the target metric and $\obsnum$.
    %     %
    %     \blt[power maps] 3) Only power maps were considered, so estimators for
    %     other classes of maps are yet to be investigated.
    %     %
    %     \blt[single frequency band] 4) All measurements took place in the same
    %     frequency band, so some conclusions may not carry over to other parts
    %     of the RF spectrum.

    % \end{bullets}

    % \blt[future work]Future work will include more extensive data
    % collection, a more systematic optimization of the parameters of
    % traditional estimators, consideration of other classes of maps, and
    % the development of further hybrid estimators that combine CNNs and
    % more traditional approaches.

\end{bullets}

\printmybibliography
\end{document}